\documentclass{aa}  

\usepackage{graphicx}

\usepackage{mathtools}
\usepackage{siunitx}
\usepackage[bottom]{footmisc}
\usepackage{tablefootnote}
\usepackage[flushleft]{threeparttable}
\usepackage{adjustbox}
\usepackage{subcaption}
\usepackage{float}

\usepackage{txfonts}
\begin{document}

   \title{Dwarf Galaxies in the MATLAS Survey: The satellite system of NGC\,474 under scrutiny with MUSE}
   \author{Oliver M\"uller
          \inst{1}    
        \and 
        Francine R. Marleau\inst{2}
          \and
          Nick Heesters\inst{1}
                    \and
          Pierre-Alain Duc\inst{3}
                    \and
          Marcel S. Pawlowski\inst{4}
          \and
           M\'elina Poulain\inst{5}
                      \and
           Rebecca Habas\inst{6}
           \and
           Elisabeth Sola\inst{7}
                      \and
           Mathias Urbano\inst{3}
                      \and
           Rory Smith\inst{8}
           \and
           Patrick Durrell\inst{9}
            \and
           Eric Emsellem\inst{10}
           \and
           Rubén Sánchez-Janssen\inst{11}
           \and
           Sungsoon Lim\inst{12}
           \and
           Sanjaya Paudel\inst{12}
          }
 \titlerunning{The satellite system of NGC\,474}
   \institute{Institute of Physics, Laboratory of Astrophysics, Ecole Polytechnique Fédérale de Lausanne (EPFL), 1290 Sauverny, Switzerland\\              
   \email{oliver.muller@epfl.ch}
               \and
    Institut f{\"u}r Astro- und Teilchenphysik, Universit{\"a}t Innsbruck, Technikerstra{\ss}e 25/8, Innsbruck, A-6020, Austria
                               \and
             Observatoire Astronomique de Strasbourg  (ObAS), Universite de Strasbourg - CNRS, UMR 7550 Strasbourg, France
             \and
                          Leibniz-Institut für Astrophysik Potsdam (AIP), An der Sternwarte 16, D-14482 Potsdam, Germany
             \and
                 Space Physics and Astronomy Research Unit, University of Oulu, P.O. Box 3000, FI-90014, Oulu, Finland
                     \and
                     INAF - Astronomical Observatory of Abruzzo, Via Maggini, 64100, Teramo, Italy
                     \and
                     Institute of Astronomy, Madingley Rd, Cambridge, CB3 0HA, UK
                     \and
                     Departamento de F\'isica, Universidad T\'ecnica Federico Santa Mar\'ia, Santiago, Chile
                              \and
                              Youngstown State University, One University Plaza, Youngstown, OH 44555 USA
\and
European Southern Observatory, Karl-Schwarzschild Straße 2, D-85748 Garching bei München, Germany
\and
UK Astronomy Technology Centre, Royal Observatory, Blackford Hill, Edinburgh, EH9 3HJ, UK
\and
                     Department of Astronomy and Center for Galaxy Evolution Research, Yonsei University, Seoul 03722
             }
   \date{Received tba; accepted tba}

 
  \abstract
   {A recent study of the distribution of dwarf galaxies in the MATLAS sample in galaxy groups  revealed an excess of flattened satellite structures, reminiscent of the co-rotating planes of dwarf galaxies discovered in the local Universe. If confirmed, this lends credence to the plane-of-satellite problem and further challenges the standard model of hierarchical structure formation. However, with only photometric data {and no confirmation of the satellite membership}, the study could not address the plane-of-satellite problem {in full detail}.  Here we present spectroscopic follow-up observations of one of the most promising planes-of-satellites candidates in the MATLAS survey, the satellite system of NGC\,474. Employing MUSE at the VLT {and full spectrum fitting}, we studied 13 dwarf galaxy candidates and confirmed nine to be members of the field around NGC\,474. {Measuring the stellar populations of all observed galaxies, we find that the MATLAS dwarfs have lower metallicities than the Local Group dwarfs at given luminosity.}
   Two dwarf galaxies may form a pair of satellites based on their close projection and common velocity. {Within the virial radius, we do not find a significant plane-of-satellites, however, there is a sub-population of six dwarf galaxies which seem to be anti-correlated in phase-space. Due to the low number of dwarf galaxies, this signal may arise by chance.} 
   With over 2000 dwarf galaxy candidates found in the MATLAS survey, {this remains} an intriguing data set to study the plane-of-satellites problem in a statistical fashion once more follow-up observations have been conducted.
   }

   \maketitle
%

\section{Introduction}
The arrangement and movement patterns of dwarf galaxies surrounding both the Milky Way and the Andromeda galaxy have ignited a continuous debate regarding their compatibility with the prevailing $\Lambda$+Cold Dark Matter ($\Lambda$CDM) standard cosmological model \citep[e.g., ][]{2005A&A...431..517K,2007MNRAS.374...16L,2013MNRAS.435.2116P,2013MNRAS.429.1502W,2014ApJ...784L...6I,2015ApJ...800...34G,2015MNRAS.452.3838C,2015MNRAS.449.2576C,2016MNRAS.460.4348B,2017MNRAS.465..641F,2021NatAs...5.1185P,2022NatAs...6..897S,2023ApJ...942...78S,2024arXiv240108143S,2024arXiv240203288G}. Numerous studies over five decades (e.g., \citealt{1976MNRAS.174..695L,2006MNRAS.365..902M,2006AJ....131.1405K,2008ApJ...680..287M,2012MNRAS.423.1109P,2013Natur.493...62I,2013ApJ...766..120C,2020ApJ...901...43S,2024A&A...681A..73T,2024arXiv240416110J}) have explored these observations, with a particular focus on the satellite systems' structure and kinematics.
Today, this debate has been dubbed the plane-of-satellites problem, and in short describes that the dwarf galaxies are arranged in thin, planar-like structures, in which they seem to co-orbit around their host galaxies (see \citealt{2018MPLA...3330004P,2021NatAs...5.1185P} for reviews). This is in stark contrast to the more pressure supported halos found in cosmological simulations.

Satellite systems are pivotal tests of $\Lambda$CDM, as their dynamics are primarily governed by gravitational forces on scales of hundreds of kiloparsecs, mostly unaffected by internal baryonic processes \citep{2018MPLA...3330004P,2018Sci...359..534M}. Consequently, numerous research teams have extended their investigations beyond the Local Group, exploring analogous structures. Beyond the Local Group, claims of flattened distributions have been proposed around several giant galaxies \citep{2013AJ....146..126C,2017A&A...602A.119M,MuellerTRGB2018,2024A&A...683A.250M,2021A&A...652A..48M,2021ApJ...917L..18P,2023MNRAS.521.4009C,2024ApJ...966..188M,2024MNRAS.528.2805K,2024arXiv240506016P}. The best studied of these is the lenticular galaxy  Cen\,A and its satellite system, which exhibits a statistically significant correlation in phase-space \citep{2015ApJ...802L..25T,Muller2016,2018Sci...359..534M,Muller2019,Muller2021b,2023MNRAS.519.6184K}. Intriguingly, these flattened structures often align with the local cosmic web, suggesting a potential common formation scenario for such configurations \citep{2015MNRAS.452.1052L,2019MNRAS.490.3786L,2024A&A...683A.250M}.

While many surveys target individual galaxies, the Mass Assembly of early-Type GaLAxies with their fine  Structures (MATLAS, \citealt{2015MNRAS.446..120D,2020MNRAS.498.2138B}) survey observed a total of 180 early type galaxies in the nearby universe{ employing MegaCam at CFHT to a limiting surface brightness magnitude of $\approx$28.5 mag/arcsec$^2$.} Over 2000 dwarf galaxies were discovered surrounding these giants \citep{2020MNRAS.491.1901H,2021MNRAS.506.5494P}, enabling the statistical analysis of their satellite systems. From a subsample of 119 satellite systems, \citet{2021A&A...654A.161H} identified a statistically significant population of 31 flattened dwarf structures, based on their on-sky positions. This analysis, however, could not take into account any velocity trend, because the necessary observations were not available at the time. It is a time consuming task to get spectroscopy of the mostly quiescent dwarf galaxies, needing 8m class telescopes and long exposures \citep[e.g., ][]{2019ApJ...874L..12D,2020A&A...640A.106M,2023A&A...676A..33H,2023MNRAS.526.4735F}, and to complete a full dwarf satellite system requires hours of dedicated observations. 

{A further issue in the analysis of the MATLAS fields was that most of the dwarf galaxy candidates did not have a confirmed membership status, meaning that the measurement of the flattening was based on the assumption that all candidates were satellites of the central galaxy. This may not be the case, as \citet{2023A&A...676A..33H} showed that only 75\% of  a subsample of 56 MATLAS dwarf galaxy candidates were confirmed to be actual members of their putative hosts through follow-up observations. This implies that the identified flattened structures in \citet{2021A&A...654A.161H} can have spurious members and the observed satellite distribution may change depending of membership confirmation/rejection.}

In this paper, we present the first follow-up observation of an almost complete satellite system in the MATLAS survey, around the iconic NGC\,474 galaxy, employing the Multi Unit Spectroscopic Explorer (MUSE) instrument mounted at the Very Large Telescope in Chile. {We analyze the dwarfs by full spectrum fitting, which yields velocities, stellar metallicities and ages,  and discuss the phase-space distribution of the satellite system in detail}.

\section{Data and Methods}
\label{data}

\begin{figure*}[ht]
    \centering
        \includegraphics[height=2.3cm]{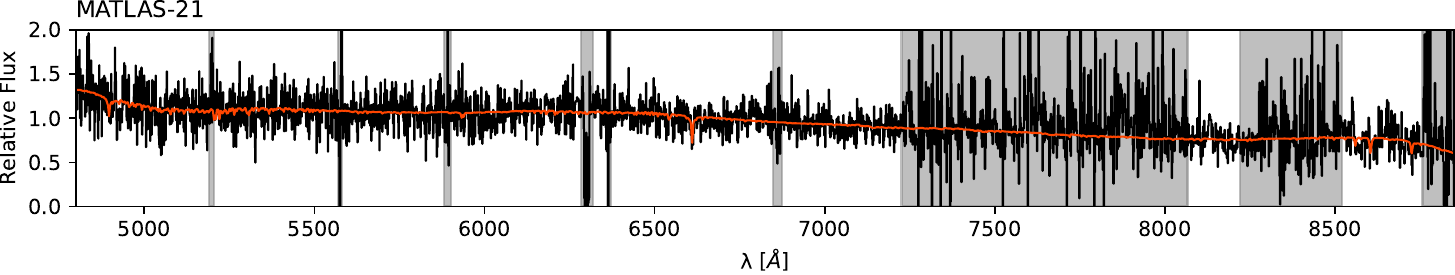}
        \includegraphics[height=2.3cm]{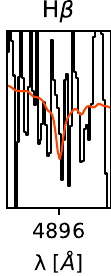}
        \includegraphics[height=2.3cm]{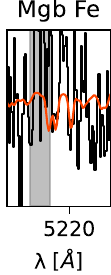}
        \includegraphics[height=2.3cm]{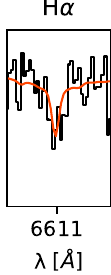}
        \includegraphics[height=2.3cm]{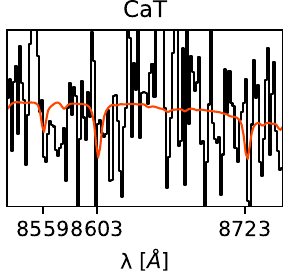}\\
        \includegraphics[height=2.3cm]{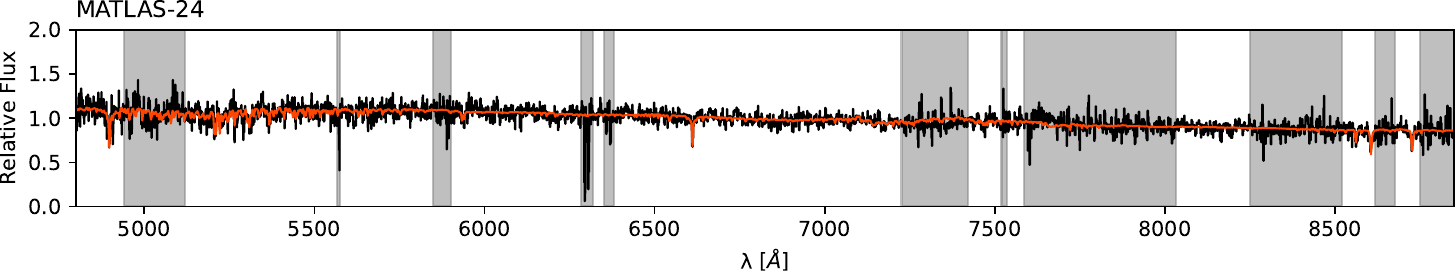}
        \includegraphics[height=2.3cm]{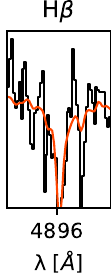}
        \includegraphics[height=2.3cm]{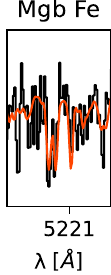}
        \includegraphics[height=2.3cm]{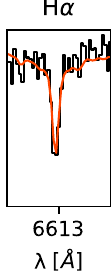}
        \includegraphics[height=2.3cm]{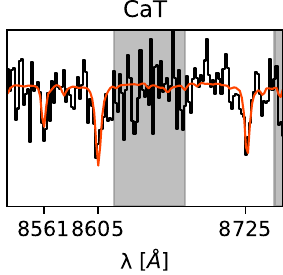}\\
        \includegraphics[height=2.3cm]{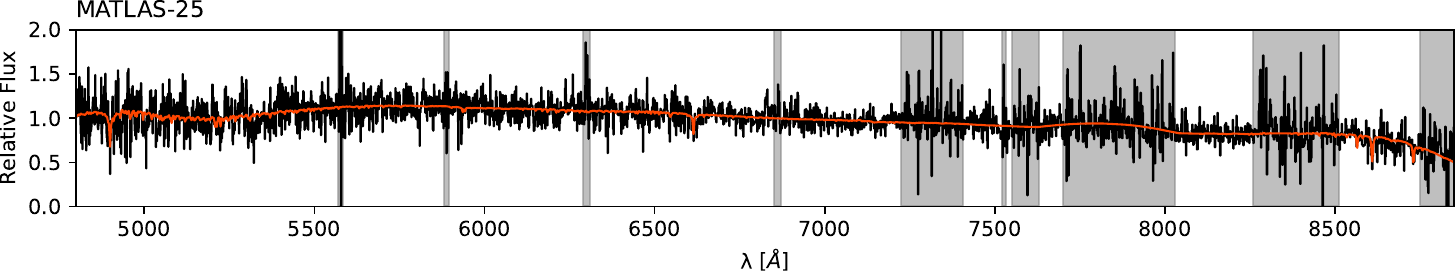}
        \includegraphics[height=2.3cm]{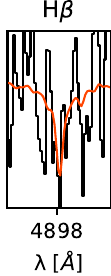}
        \includegraphics[height=2.3cm]{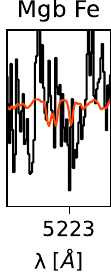}
        \includegraphics[height=2.3cm]{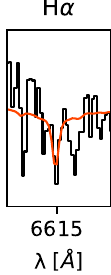}
        \includegraphics[height=2.3cm]{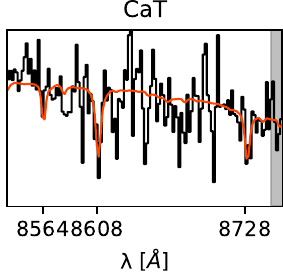}\\
        \includegraphics[height=2.3cm]{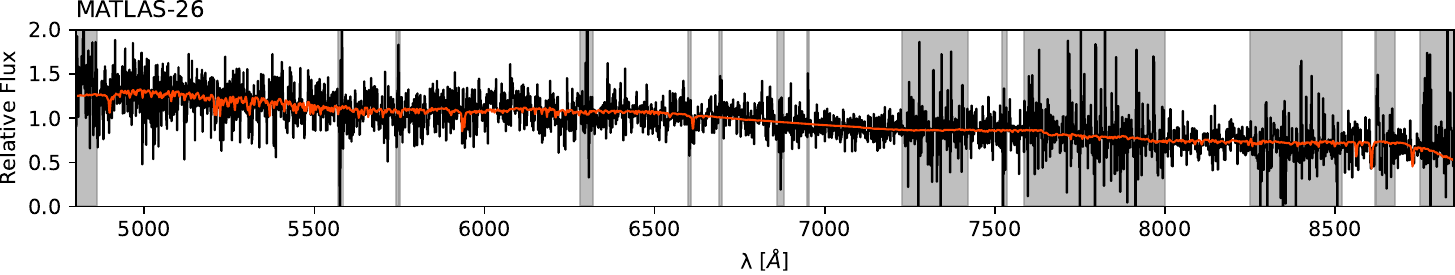}
        \includegraphics[height=2.3cm]{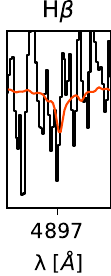}
        \includegraphics[height=2.3cm]{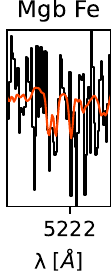}
        \includegraphics[height=2.3cm]{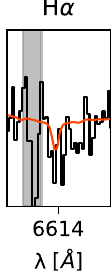}
        \includegraphics[height=2.3cm]{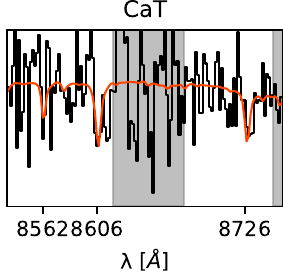}\\
            \includegraphics[height=2.3cm]{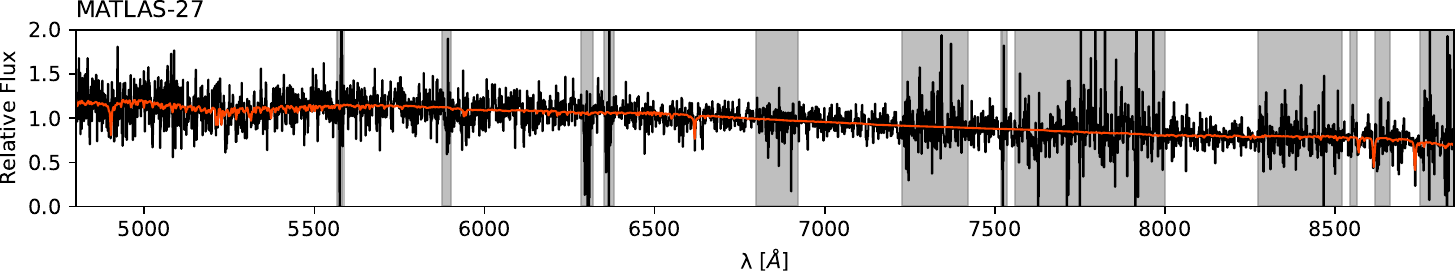}
        \includegraphics[height=2.3cm]{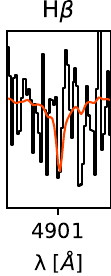}
        \includegraphics[height=2.3cm]{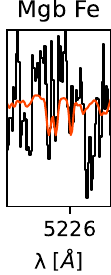}
        \includegraphics[height=2.3cm]{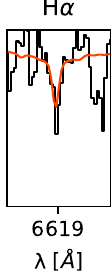}
        \includegraphics[height=2.3cm]{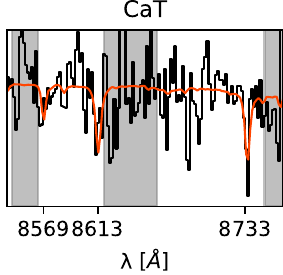}\\
        \includegraphics[height=2.3cm]{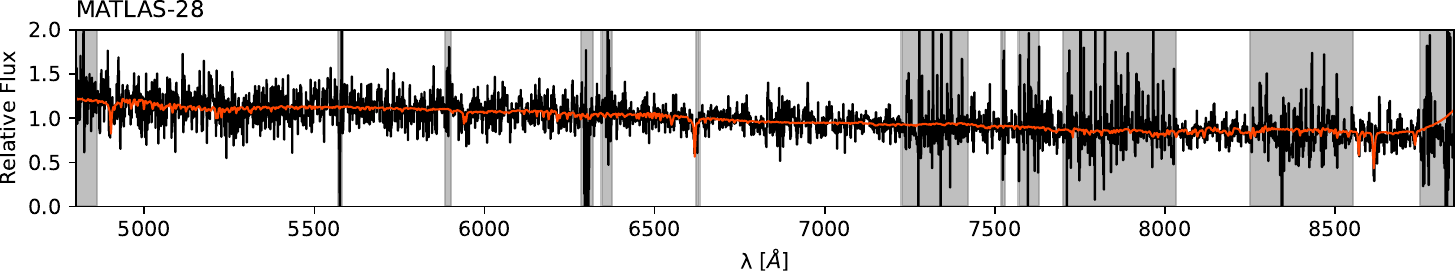}
        \includegraphics[height=2.3cm]{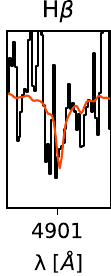}
        \includegraphics[height=2.3cm]{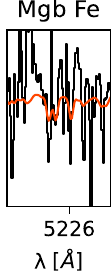}
        \includegraphics[height=2.3cm]{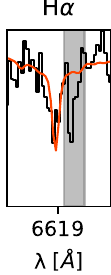}
        \includegraphics[height=2.3cm]{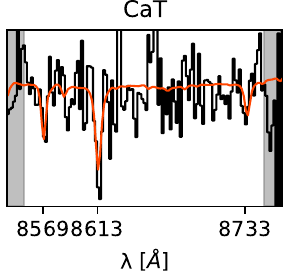}\\
    \caption{The spectra of the confirmed dwarf galaxies in the field around NGC\,474. The spectra (black line) and the best fit from pPXF (red line) is plotted in the left panel over the full spectral coverage of MUSE. The gray areas indicate masked regions. {The right panels highlight features of interest that were relevant for the fitting.}
    }
    \label{fig:spectra}
\end{figure*}

\begin{figure*}[ht]
    \centering
        \includegraphics[height=2.3cm]{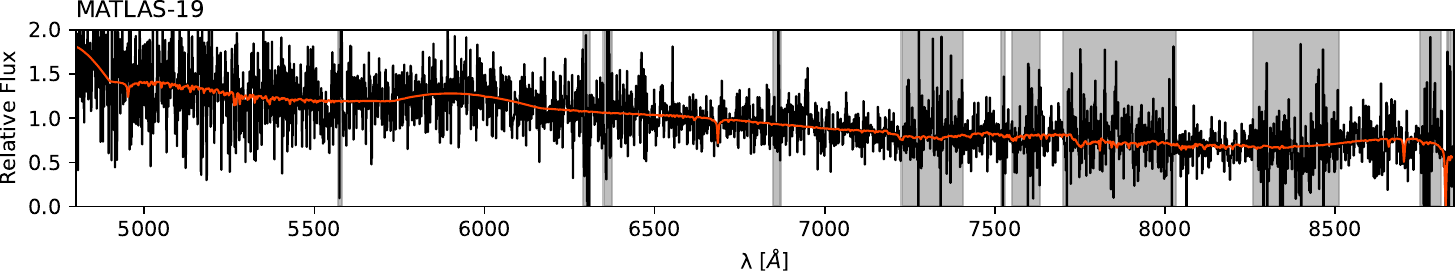}
        \includegraphics[height=2.3cm]{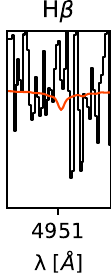}
        \includegraphics[height=2.3cm]{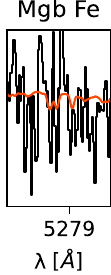}
        \includegraphics[height=2.3cm]{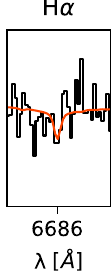}
        \includegraphics[height=2.3cm]{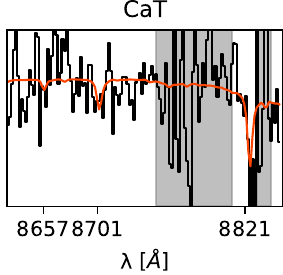}\\
        \includegraphics[height=2.3cm]{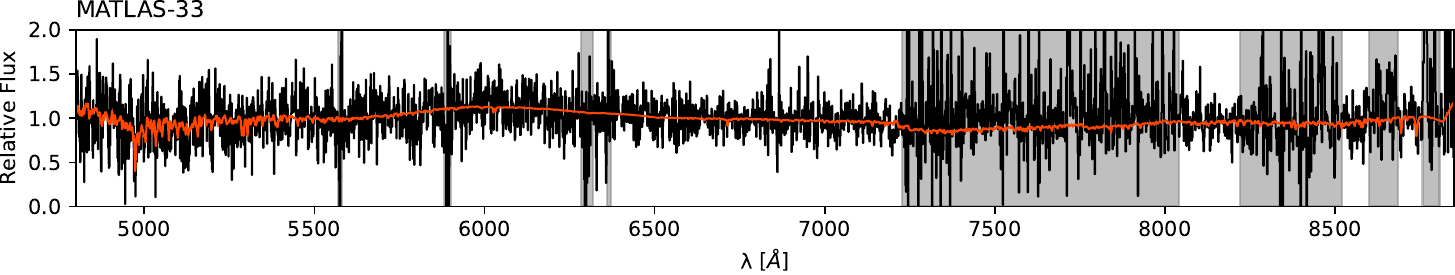}
        \includegraphics[height=2.3cm]{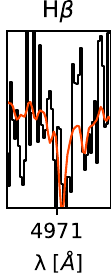}
        \includegraphics[height=2.3cm]{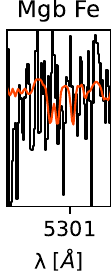}
        \includegraphics[height=2.3cm]{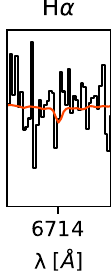}
        \includegraphics[height=2.3cm]{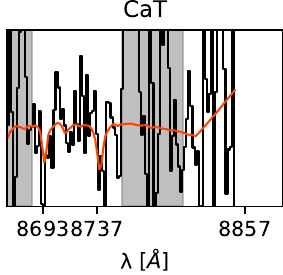}\\
        \includegraphics[height=2.3cm]{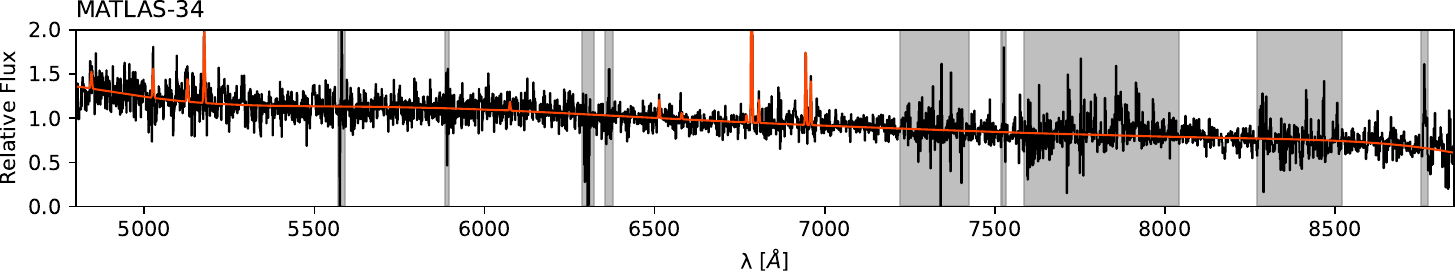}
        \includegraphics[height=2.3cm]{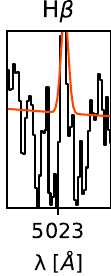}
        \includegraphics[height=2.3cm]{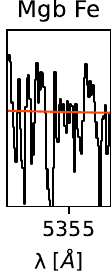}
        \includegraphics[height=2.3cm]{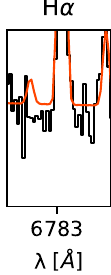}
        \includegraphics[height=2.3cm]{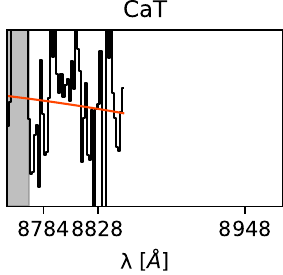}\\
    \caption{Same as Fig\,\ref{fig:spectra} but for the background galaxies.  
    }
    \label{fig:spectra_bg}
\end{figure*}

\begin{figure}[ht]
    \centering
        \includegraphics[width=0.32\linewidth]{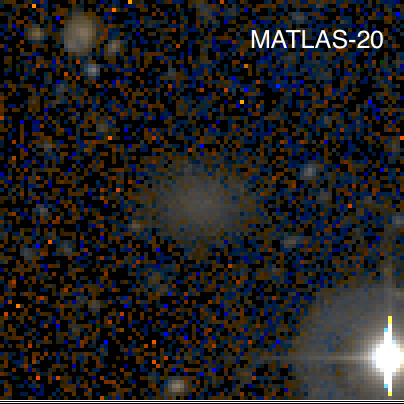}
        \includegraphics[width=0.32\linewidth]{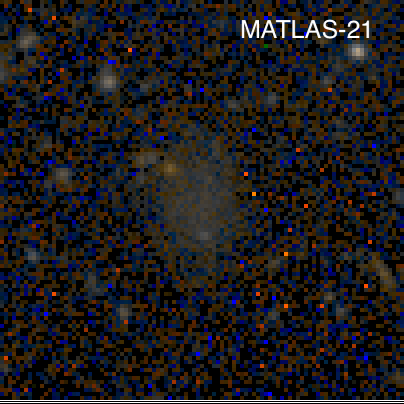}
        \includegraphics[width=0.32\linewidth]{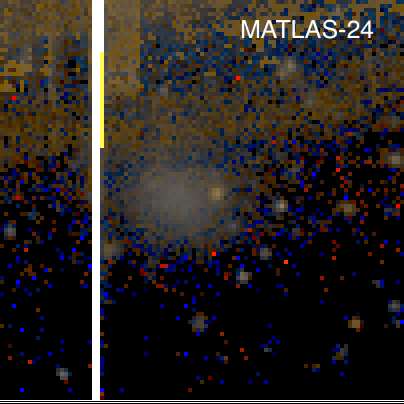}\\
        \includegraphics[width=0.32\linewidth]{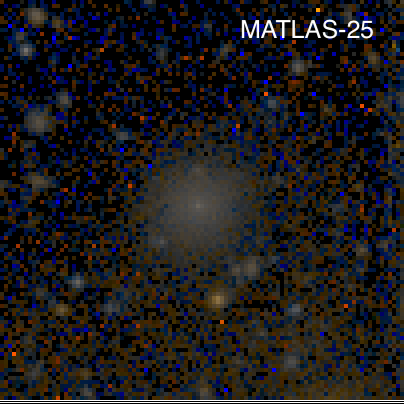}
        \includegraphics[width=0.32\linewidth]{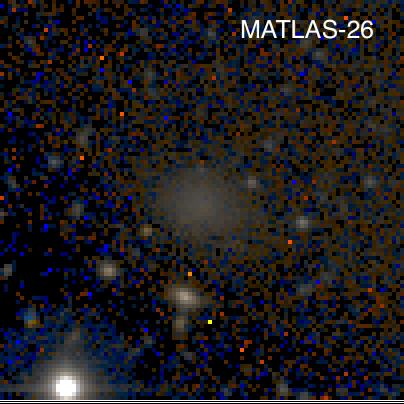}
        \includegraphics[width=0.32\linewidth]{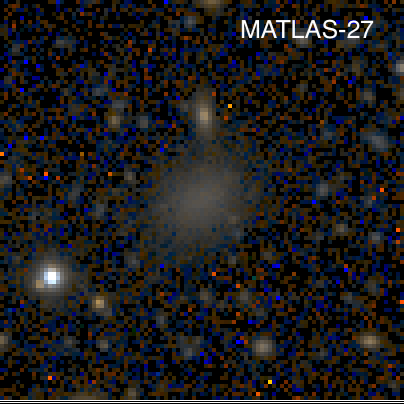}\\
        \includegraphics[width=0.32\linewidth]{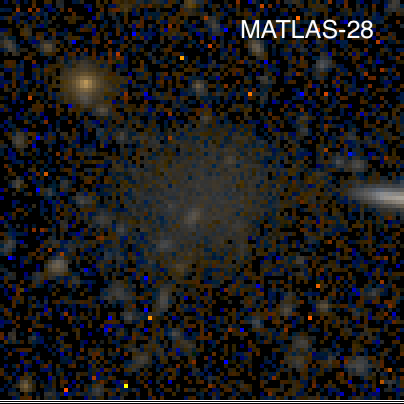}
        \includegraphics[width=0.32\linewidth]{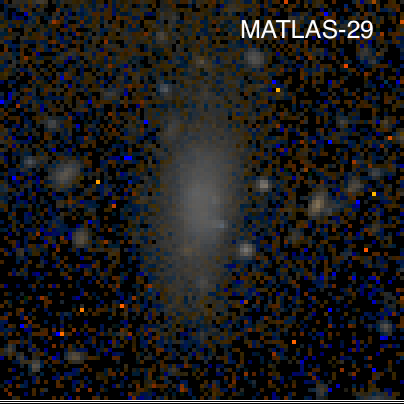}
        \includegraphics[width=0.32\linewidth]{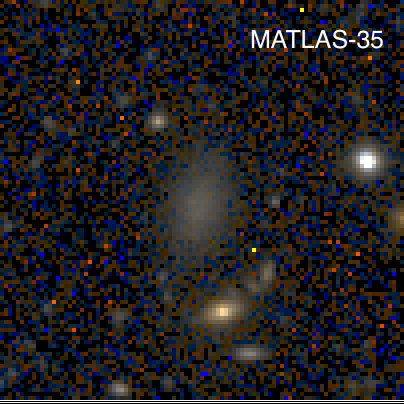}
    \caption[Caption for LOF]{Color-composite of the confirmed dwarf galaxies {in the field surrounding} NGC\,474 from the MATLAS survey, processed from the MegaCam images and downloaded from the MATLAS webpage\protect\footnotemark. North to the top, East to the left. One side is one arcmin.} 
    \label{fig:snapshopts_confirmed}
\end{figure}
\footnotetext{see \url{http://obas-matlas.u-strasbg.fr/WP/}, last accessed 30.01.2024}

\begin{figure}[ht]
    \includegraphics[width=0.32\linewidth]{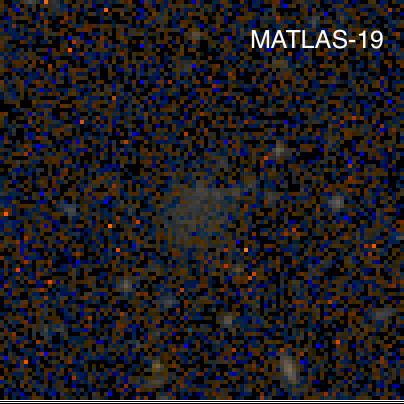}
        \includegraphics[width=0.32\linewidth]{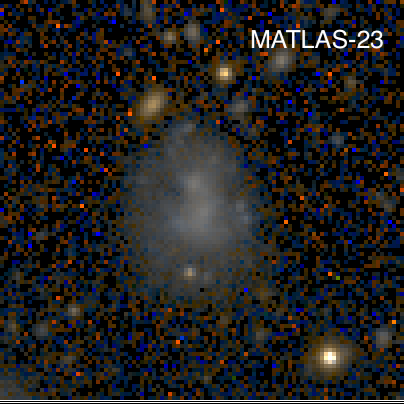}
        \includegraphics[width=0.32\linewidth]{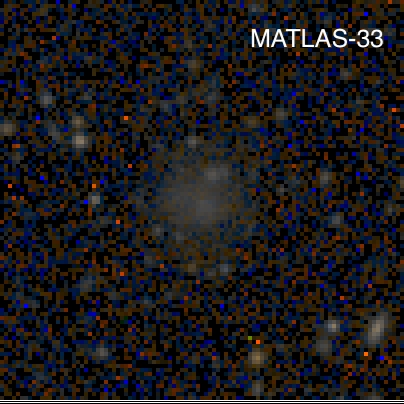}\\
        \includegraphics[width=0.32\linewidth]{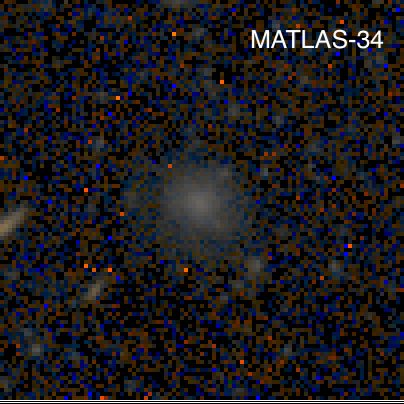}
    \caption[Caption for LOF]{Color-composite of the rejected dwarf galaxies of NGC\,474 from the MATLAS survey, processed from the MegaCam images and downloaded from the MATLAS webpage.  North to the top, East to the left. One side is one arcmin.} 
    \label{fig:snapshopts_unconfirmed}
\end{figure}

The NGC\,474 group was selected as  a follow up target because it was identified to host the most significant flattened distribution of dwarf galaxy candidates \citep{2021A&A...654A.161H} observable in the South during the semester P111.
The data was acquired through the  European Southern Observatory (ESO) programs 109.22ZV, 110.23QF.001, and 111.24KF (PI: Marleau) using the Multi Unit Spectroscopic Explorer \citep[MUSE;][]{Bacon2010,2012Msngr.147....4B} at UT4 of the Very Large Telescope (VLT) on Cerro Paranal, Chile. MUSE is an integral field unit (IFU) with an $1\times1$\,arcmin$^2$ field of view, a spatial sampling of 0.2\,arcsec/pixel, a nominal wavelength range of 480-930\,nm, and a resolving power between 1740 (480 nm) and 3450 (930 nm). In total we observed 13 dwarf galaxy candidates from the MATLAS survey. We include three dwarf galaxies -- MATLAS-20, MATLAS-29, and MATLAS-35 -- that were already analyzed and presented in \citet{2023A&A...676A..33H}.  In order to observe the major part of the galaxy (approx. to two effective radii) but have  more clean sky, we applied an offset of the order of $\pm$10 arcsec in RA and Dec from the center of the MUSE field of view.
The MUSE data products are available through the ESO Science Archive, which includes the calibrated data run through the standard MUSE pipeline \citep{Weilbacher2012,2020arXiv200608638W}. This included the bias and the flat-field correction, the astrometrical calibration,  the sky-subtraction, and the wavelength and flux calibration \citep{Hanuschik2017}\footnote{see also \url{http://www.eso.org/observing/dfo/quality/PHOENIX/MUSE/processing.html}}. 

For the sky subtraction, we directly used the data cube because the target dwarf galaxies are small enough that they leave enough space to estimate the sky. No sky exposures were taken. \citet{Muller2021a} estimated that with 20\% of the MUSE field being free for the sky subtraction, better results are achieved when employing the science data cube instead of a sky offset.
To further reduce the sky residual lines, we employed the Zurich Atmosphere Purge (ZAP) principal component analysis algorithm \citep{Soto2016}. 
To select the empty sky patches on the data we ran the python implementation of Source Extractor \citep{SExtractor} called SEP \citep{SEP} with a sigma threshold of 0.5, which creates a segmented fits file with all {detected} 
sources. This segmentation map {served as} mask for ZAP. We extracted the integrated spectra {using} 
an elliptical aperture, which was adjusted on the the collapsed cubes. We again used SEP to mask foreground stars and background objects with different thresholds in a trial-by-error fashion. Additionally, we masked objects by hand where needed.

To extract the line-of-sight velocities and stellar population properties, we employed the Python implementation of the Penalized PiXel-Fitting (pPXF, \citealt{2004PASP..116..138C,2017MNRAS.466..798C}) algorithm. We followed the same steps as described in several previous studies employing MUSE and pPXF for this task \citep{2019A&A...625A..76E,2019A&A...625A..77F,2020A&A...640A.106M,2020A&A...634A..53F,2022A&A...667A.101F,2023A&A...676A..33H}. 
In summary, our approach involves utilizing a set of Single Stellar Population (SSP) spectra sourced from the eMILES library \citep{2016MNRAS.463.3409V}, encompassing metallicities {[M/H]} spanning from solar levels down to $-$2.27 dex and ages ranging from 70 Myr to 14.0 Gyr. We adopt a Kroupa initial mass function (IMF, \citealt{2001MNRAS.322..231K}).
The SSP library spectra undergo convolution with the line-spread function, following the methodology outlined in \citet{2017A&A...608A...5G} and detailed in the Appendix of \citet{2019A&A...625A..76E}. A variance spectrum is derived from the masked data cube and incorporated into pPXF to enhance the fitting process.
For the kinematic fit, we employ 8 degrees of freedom for the multiplicative polynomial and 12 degrees for the additive polynomial \citep{2019A&A...625A..76E}. In the age and metallicity fits, we constrain the velocity, omit additive polynomials, and maintain the 12th degree in the multiplicative polynomial  \citep{2019A&A...625A..77F}. We utilize pPXF weights to compute mean metallicities, mean ages, and stellar mass-to-light ratios from the SSP models for each galaxy. 
To enhance the fits, we mask residual sky lines not removed by ZAP. The spectra and the best-fit models of the confirmed NGC\,474 members and background objects are presented in Fig\,\ref{fig:spectra} and Fig.\,\ref{fig:spectra_bg}, and the derived properties are compiled in Table\,\ref{tab:sample} and Table\,\ref{tab:sample_bg}.

The uncertainties on the best-fit parameters are estimated using a Monte Carlo method, where residuals are reshuffled in a bootstrap approach. The uncertainties are given as the 1$\sigma$ standard deviation of the posterior distribution. 
We calculate the S/N ratio per pixel in a region between 6600 and 6800 Å, devoid of strong absorption or emission lines. This ratio is determined as the mean fraction between flux and the square root of the variance, with the latter multiplied by the $\chi^2$ value estimated by pPXF. Because the main goal of the observations was to measure the line-of-sight velocities, the targeted S/N ratio was 5. Therefore, we advise that the age and metallicity estimations to be taken with a precaution (e.g. \citealt{2019A&A...628A..92F} argue that one needs a S/N ratio of $>$10 to determine a metallicity within 0.2\,dex).

\section{The mass-metallicity relation}
\begin{figure}[ht]
    \centering
    \includegraphics[width=\linewidth]{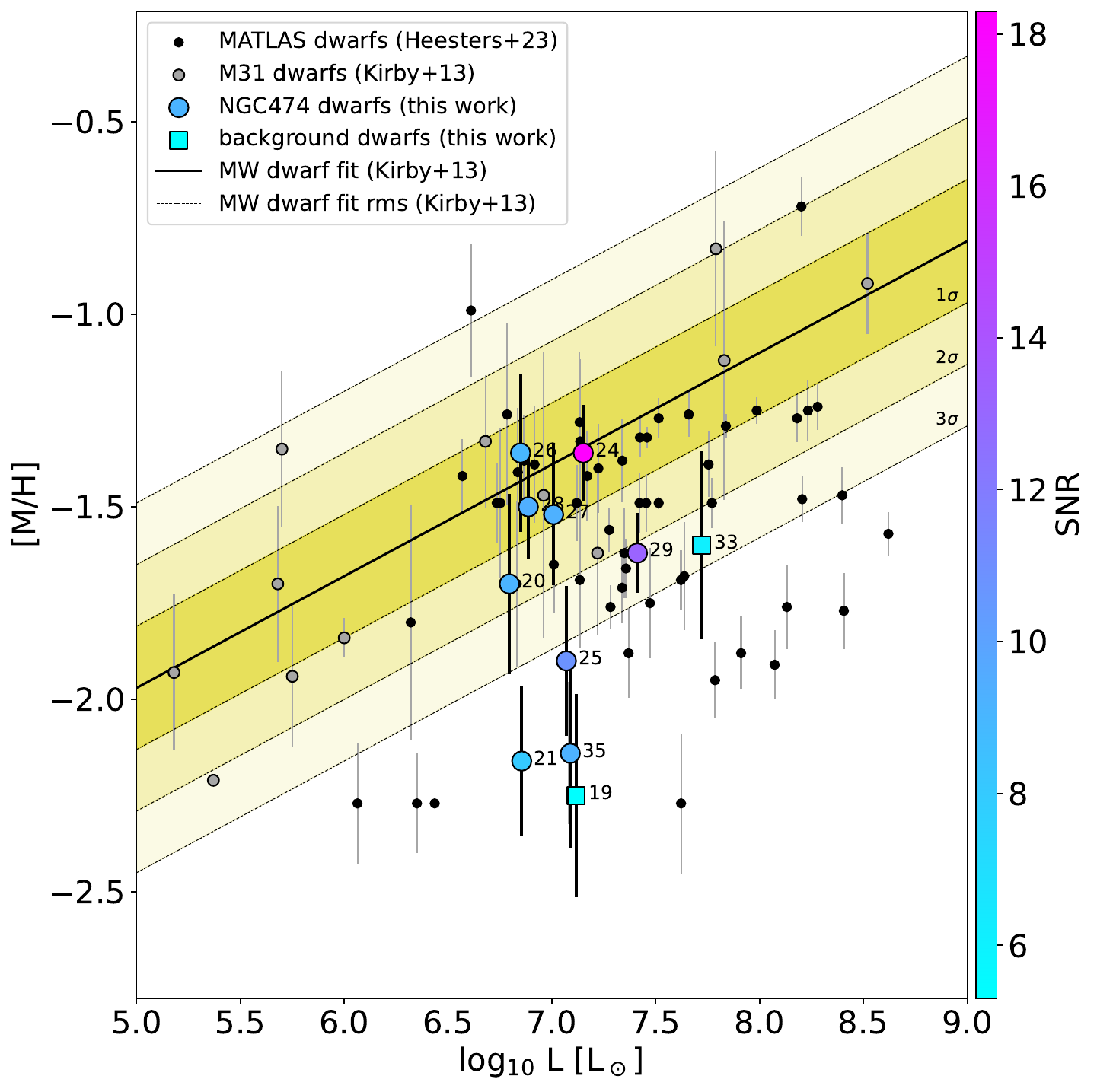}
    \caption{{The mass-metallicity relation for a reference sample of dwarf galaxies from the MATLAS survey \citep{2023A&A...676A..33H}, {the M31 dwarfs \citep{2013ApJ...779..102K}, and the dwarf galaxies presented here. The black line corresponds to Eq. 3 from \citet{2013ApJ...779..102K} and the  yellow regions indicate the 1,2, and 3\,$\sigma$ intervals.}}}
    \label{fig:mass_metal}
\end{figure}

{
The dwarf galaxies in the Local Group seem to follow a well defined relation in metallicity and mass \citep{2013ApJ...779..102K}. \citet{2023A&A...676A..33H} observed 56 dwarf galaxies from the MATLAS survey \citep{2020MNRAS.491.1901H,2021MNRAS.506.5494P} in different fields with MUSE using a similar data reduction pipeline as presented here\footnote{{The main difference between the pipeline used here and from \citet{2023A&A...676A..33H} is the extraction of the sky pixels using different tools, namely MTObject \citep{teeninga2015improved} or Source Extractor. This caused a difference in the stellar population properties for MATLAS-2019 between \citet{2023A&A...676A..33H} and \citet{2020A&A...640A.106M}, which was pointed out by \citet{2024MNRAS.529.3210B}.}}. They found an offset from this relation towards more metal-poor stellar populations, which they argue may be due to systematic differences in the estimation of the metallicity (coming from mainly CaT fitting of individual stars for the Local Group or full spectrum fitting for the whole galaxy for the MATLAS dwarfs). 
This begs the question whether we find a similar systematic difference with our new data.
For that we show the mass-metallicity relation in Fig.\,\ref{fig:mass_metal} and compare it to the larger MATLAS dwarf sample and the Local Group dwarfs.  Similar to \citet{2023A&A...676A..33H}, we find that in general the dwarfs from the MATLAS sample seem to be more metal poor than the Local Group dwarfs. We find four MATLAS dwarfs (MATLAS-19, MATLAS-21, MATLAS-25, and MATLAS-35) from our sample outside of the 3$\sigma$ bounds defined by \citet{2013ApJ...779..102K}. This is not solely driven by the S/N ratio, because three of these dwarfs have a S/N ratio of $\approx$9-10, and only one has a low ratio of 5.  The dwarfs studied here follow the same trend that was found in the previous MATLAS MUSE observations \citep{2023A&A...676A..33H} and may indicate a deviation from the mass-metallicity relation. Whether this is coming from differences in the measurement of the metallicity or rather has a physical nature remains to be seen. 
}

\section{The NGC\,474 satellite system}

Before we dive into the phase-space analysis of the NGC\,474 system, we need a clear picture of the membership and satellite nature of the dwarf galaxies. But let us first assess the giant galaxies in the field.

\begin{figure}[ht]
    \centering
    \includegraphics[width=\linewidth]{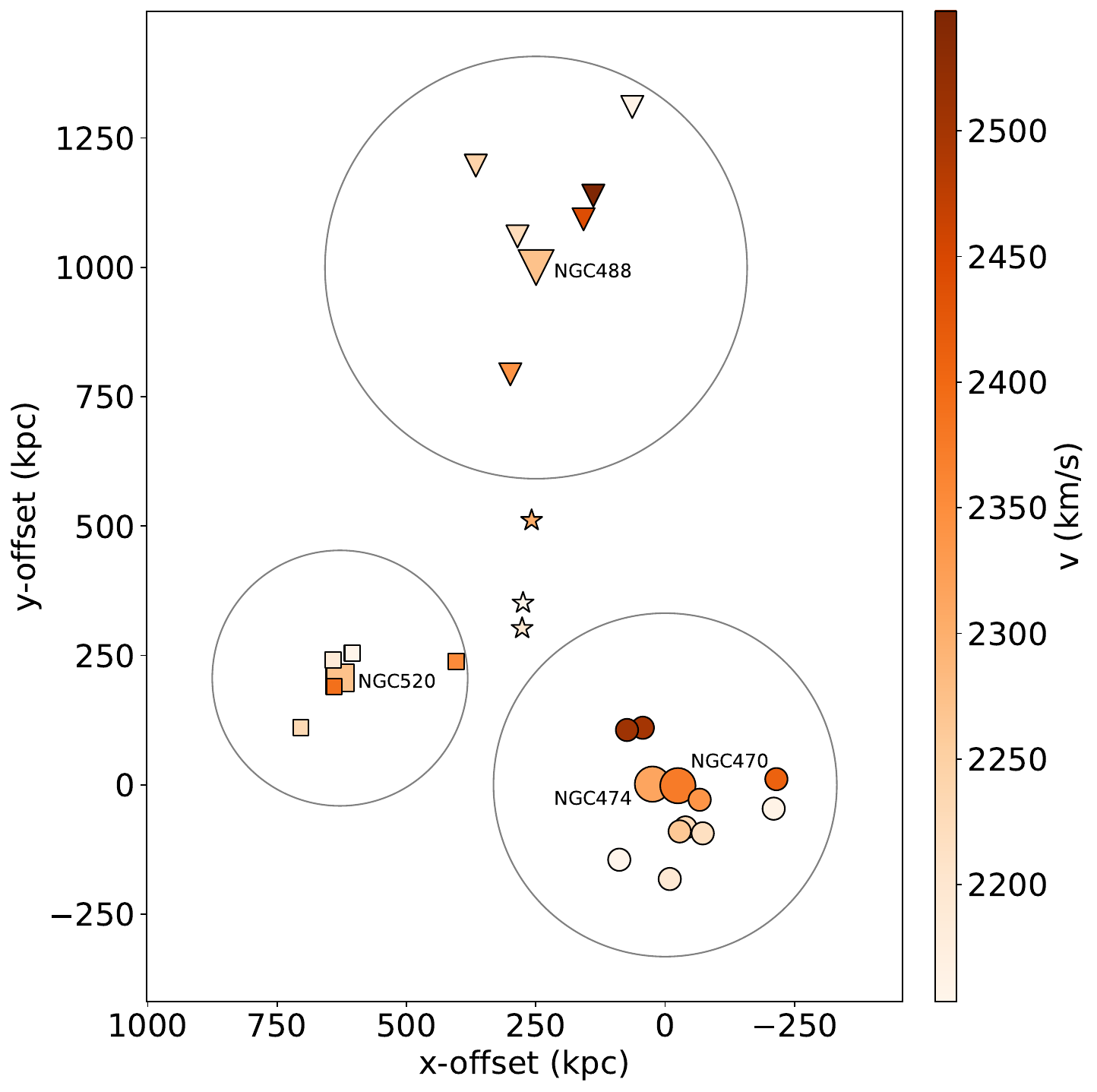}
    \caption{The larger field in XY coordinates at a distance of 31\,Mpc. The X-axis follows the RA-axis, with positive numbers towards East, the Y-axis follows the Dec-axis, with positive numbers towards North.
    Colors indicate the line-of-sight velocities. The circles correspond to the NGC\,474/NGC470 system, the squares to the NGC\,520 system, and the triangles to the NGC\,488 system. The stars indicate field dwarfs (being outside of the virial radius). The virial radii of the systems are indicated with gray circles.
    }
    \label{fig:field_large}
\end{figure}

The NGC474 system includes the post-merger galaxy NGC\,474 and close-by the spiral galaxy NGC470 with a separation of $\sim$5.3 arcmin. 
Both have similar stellar masses  with 6-9$\times$10$^{10}$\,M$_\odot$ \citep{2010PASP..122.1397S,2013MNRAS.432.1709C} and 8$\times$10$^{10}$\,M$_\odot$ \citep{2010PASP..122.1397S}, respectively.
In the further neighborhood, two other giant galaxies reside, namely NGC\,488 and NGC\,520, with separations of $\sim$1.9 deg and $\sim$1.2 deg with respect to NGC\,474, respectively (see Fig.\,\ref{fig:field_large}). At an adopted distance of 31\,Mpc \citep{2011MNRAS.413..813C} for NGC\,474\footnote{We note that several distance estimates exist based on different methods and calibrations \citep{2007ApJ...668..130C,2011MNRAS.413..813C,2013AJ....146...86T}, ranging between 28 and 33\,Mpc. For consistency reason with other publications using the MATLAS data, we use the values provided by \citet{2011MNRAS.413..813C}.}, this translates into physical, projected separations of $\sim$50\,kpc, $\sim$1000\,kpc, and $\sim$650\,kpc for NGC\,470, NGC\,488, and NGC\,520, respectively. All galaxies share similar line-of-sight velocities, namely 2375\,km/s for NGC\,470 \citep{2005ApJS..160..149S}, 2315\,km/s for NGC\,474 \citep{2011MNRAS.413..813C}, 2272\,km/s for NGC\,488 \citep{1993ApJS...88..383L}, and 2281\,km/s for NGC\,520 \citep{1993ApJS...88..383L}. Errors are between 1 and 5\,km/s. The closeness in velocity space will make it difficult to disentangle membership of the dwarf galaxies based on their velocities alone. 
There are also two background galaxies within the projected virial radius of NGC\,474, NGC\,467 with a velocity of $\sim$5500 km/s \citep{1999ApJS..121..287H}, and NGC\,479 with a velocity of $\sim$5300 km/s \citep{2022ApJS..261...21Y}. With such large differences in velocities to NGC\,474, there is no confusion between their dwarf satellite members.

The MATLAS survey covered a 1.044$\times$1.153 deg$^2$ field around NGC\,474, which covers most of the virial radius  (330\,kpc, or $\approx$0.6\,deg).  {We calculated the virial radius of the early-type galaxy NGC474 using the abundance matching relation between the effective radius and $R_{200}$ \citep{2013ApJ...764L..31K}, with an effective radius of 33\,kpc \citep{2013MNRAS.432.1709C}.  For the  late-type galaxies shown in Fig.\ref{fig:field_large} we derive the virial radii from the baryonic masses and using Eq. 11 from \citet{2012AJ....143...40M}, with a Hubble constant value of 70\,km/s/Mpc. This calculation follows the baryonic Tully-Fisher relation {as put forward in \citet{2012AJ....143...40M}}. We use an overdensity of $\Delta=200$ above the cosmic density as definition for the virial mass/radius.}

In the NGC\,474 field there were 14 dwarf galaxy candidates listed in \citet{2021MNRAS.506.5494P}, Table\,1. 
{ Additionally, there are two dwarf galaxies with known velocities, LEDA\,4765 with $v=2235\pm31$\,km/s, 
and LEDA\,1244578 with $v=2172\pm16$\,km/s \citep{2003AJ....126.2152R} that are not in the MATLAS sample.
}
If we are assuming that satellites have systemic velocities within $\pm$300\,km/s of their host galaxies \citep{2020MNRAS.491.1901H}, {every dwarf galaxy candidate with a line-of-sight velocity between 2050 and 2650 km/s may potentially be a satellite member of the NGC\,474 system. The central velocity of 2350 km/s is derived as a luminosity weighted mean between NGC\,474 and NGC\,470.} We find that this is the case for nine of the observed dwarf galaxies, see Fig.\,\ref{fig:snapshopts_confirmed}. {Four dwarf candidates (MATLAS-19, MATLAS-23, MATLAS-33, MATLAS-34, see Fig.\,\ref{fig:snapshopts_unconfirmed}) are actually  background objects}, we thus remove them from the satellite list\footnote{{MATLAS-19 and MATLAS-33 can be considered ultra-diffuse galaxies at their Hubble distance, with effective radii of 1.8\,kpc, and 2.3\,kpc, respectively.}}. {See Table\,\ref{tab:sample_bg} for their velocities.} This is consistent with the previous follow-up observations of MATLAS satellites, {where a confirmation rate of 75\% was found \citep{2023A&A...676A..33H}. Here we report a confirmation rate of 70\%.} One dwarf galaxy candidate -- MATLAS-22 - was not observed. We keep it in the list as a potential dwarf galaxy. {Another dwarf -- MATLAS-35 -- is outside of the virial radius of NGC\,474 and likely a field dwarf.}
This gives us a list of 10 confirmed dwarf satellite galaxies {and one candidate} in the NGC\,474 system.

 Two of the dwarf galaxies -- MATLAS-27 and MATLAS-28 -- share a very similar velocity (2500$\pm$11\,km/s and 2508$\pm$8\,km/s) and are close in projection with a separation of 3.45\,arcmin (or 31\,kpc at 31\,Mpc). They may be a gravitationally bound pair of satellites. Are they massive enough to be physically bound? \citet{2010ApJ...711..361G} defined a criteria for a pair of satellites to be bound when  
 \begin{equation}
      b \equiv 2GM_\mathrm{pair}/\Delta r\Delta v^2 > 1,
 \end{equation}

 where  $\Delta r$ is the physical separation between the objects, $\Delta v$ is their total velocity difference,  $M_\mathrm{pair}$ is the total mass of the pair, and  $G$ is the gravitational constant. Because we do not have the total mass of the system, we can rephrase the question to how massive the system must be to be physically bound. The two dwarfs have stellar luminosities of  $1.08\times10^7$\,L$_\odot$ and  $0.83\times10^6$\,L$_\odot$, respectively \citep{2021MNRAS.506.5494P}.  This translates into masses of $1.6\times10^7$\,M$_\odot$ for both using the measured stellar M/L ratio. If we solve for $M_\mathrm{pair}$ and using $\Delta r=31$\,kpc and $\Delta v=8$\,km/s we get a minimal total mass of the system of $2.3 \times 10^8$\,M$_\odot$. The criteria $b>1$ is satisfied by {a total dark matter halo 7 times more massive than the stellar mass of the two combined (i.e. each dwarf has e.g. a mass-to-light ratio of 4)}. However, there are two problems: a) the velocities have an uncertainty of $\pm11$\,km/s which still allows for a large $\Delta v$, and b) the true physical separation is unknown.  {A further assumption is that the tangential velocity vectors of the two dwarfs are similar or negligible.} For b) we can assume that the 3D separation is rather $\sqrt{2}$ times the projected separation (i.e. the uncertainty in depth is equal to the projected separation). If we assume $\Delta v=16$\,km/s {(the quadratic sum of the two uncertainties and the absolute measured velocity difference)} and  $\Delta r=44$\,kpc we estimate a conservative total mass to be bound of  $1.3\times10^9$\,M$_\odot$. This is 41 times more massive than the {combined} stellar mass {of the two}. This would assume that the dwarfs have each a dark matter halo 20 times more massive than their stellar body, which for dwarf satellites in this luminosity range is reasonable \citep[e.g., ][and references therein]{2012AJ....144....4M}. We thus argue that MATLAS-27 and MATLAS-28 {could be} a physically bound pair of satellites. {Note that a dwarf galaxy satellite at this luminosity range typically resides in a stripped dark matter halo of the order of $10^{9}$ to $10^{10}$\,M$_\odot$ according to hydrodynamical simulations  \citep{2022NatAs...6..897S}.}

\begin{table*}[ht]
\caption{The confirmed NGC474 dwarf galaxy system.}             
\centering                          
\begin{tabular}{l c c c l c r c r r}        
\hline\hline                 
 Name &        RA &      Dec &  Exp. time &         vel &        [M/H] &        age &         M/L & S/N & M$_V$ \\    
 & deg & deg & s & km/s & dex & Myr & M$_\odot$ /L$_\odot$ & & mag \\    
  (1) &        (2) &      (3) &  (4) &         (5) &        (6)&        (7) &        (8) & (9) & (10) \\    
\hline      \\[-2mm]                  
MATLAS-20$^{(a)}$ & 19.584400 & 3.433197& 2628 & 2409.0$\pm$14.6 & $-$1.70$\pm$0.23 & 12.6$\pm$1.5 &  2.2$\pm$0.2 & 9.1 & $-$12.2\\
MATLAS-21 & 19.594114 & 3.327721  &  5256 & 2161.9$\pm$24.4 & $-$2.16$\pm$0.19 & 13.4$\pm$2.1 & 2.0$\pm$0.2 &  8.0 & $-$12.3 \\
MATLAS-24 & 19.848000 & 3.239670& 2628 &  2216.8$\pm$5.1 & $-$1.36$\pm$0.12 &  8.4$\pm$1.9 &  1.6$\pm$0.4 & 18.3 & $-$13.0 \\
MATLAS-25 & 19.858900 & 3.359683& 2628 & 2339.6$\pm$13.6 & $-$1.90$\pm$0.19 & 13.7$\pm$2.6 & 1.8$\pm$0.4 & 10.8 & $-$12.8 \\
MATLAS-26 & 19.930422 & 3.246573&2628 & 2263.8$\pm$12.6 & $-$1.36$\pm$0.20 & 11.7$\pm$2.1 &  2.0$\pm$0.3 &  9.0 & $-$12.3 \\
MATLAS-27 & 20.062317 & 3.616693&2628 & 2500.1$\pm$11.3 & $-$1.52$\pm$0.18 &  8.2$\pm$2.3 &  1.5$\pm$0.3 &  9.3 &  $-$12.7 \\
MATLAS-28 & 20.119040 & 3.608794 & 5913 & 2508.2$\pm$8.0 &  $-$1.5$\pm$0.13 & 10.5$\pm$2.3 &   1.8$\pm$0.3 &  9.4 &  $-$12.4 \\
MATLAS-29$^{(a)}$ & 20.146580 & 3.145582& 2628 & 2153.4$\pm$10.4 & $-$1.62$\pm$0.10 &  9.2$\pm$2.0 &  1.9$\pm$0.5 & 13.2 & $-$13.7 \\
MATLAS-35$^{(a)}$ & 20.494549 & 3.971204& 2628 & 2193.8$\pm$18.6 & $-$2.14$\pm$0.24 &  6.4$\pm$3.8 &  1.2$\pm$0.5 & 9.3 &  $-$12.9 \\
\\
LEDA\,4765$^{(b)}$ & 19.909685 & 3.261206 & --- & 2235$\pm$31  & --- &  --- & --- &   --- &  \\
LEDA\,1244578$^{(b)}$ & 19.965508 & 3.077404 & --- & 2172$\pm$16  & --- &  --- & --- &   --- & -16.0  \\

\hline
\end{tabular}
\tablefoot{ The reference for the values presented in this row are from (a): \citet{2023A&A...676A..33H}, and (b): \citet{2003AJ....126.2152R}.
Some comments on the different columns: (4) the exposure time is the total integrated exposure time on target and may include several observing blocks. (8) the M/L ratio is the stellar mass-to-light ratio derived from the pPXF fitting of the spectrum. (9) the signal-to-noise is measured on the spectrum in a region between 6600 and 6800 Å. (10) {Extinction corrected \citep{2011ApJ...737..103S} absolute magnitudes, estimated from the $g$ and $r$ band magnitudes of  \citet{2021MNRAS.506.5494P} for the MATLAS dwarfs, and \citet{2018MNRAS.474.1873W} for LEDA\,1244578}.}
\label{tab:sample}
\end{table*}

\begin{table*}[ht]
\caption{The rejected NGC\,474 dwarf satellite candidates.}             
\centering                          
\begin{tabular}{l c c c l c r c c c}        
\hline\hline                 
 Name &        RA &      Dec &  Exp. time &         vel &        [M/H] &        age &         M/L & S/N & M$_V$ \\    
 & deg & deg & s & km/s & dex & Myr & M$_\odot$ /L$_\odot$ & &mag \\    
  (1) &        (2) &      (3) &  (4) &         (5) &        (6)&        (7) &        (8) & (9) & (10) \\    
\hline      \\[-2mm]                  
MATLAS-19 & 19.558893 & 2.903885 & 5256 & 5576.0$\pm$39.2 &  $-$2.25$\pm$0.26 & 5.8$\pm$2.0  &  1.12$\pm$0.25&   5.3  & $-$13.0\\
MATLAS-23$^{(c)}$ & 19.711599 &	3.561740 & --- & 4934$\pm$8 & --- & --- & ---  & --- & $-$15.4 \\
MATLAS-33 & 20.384933 &	3.671406 & 2628 &  6844.8$\pm$19.6 & $-$1.60$\pm$0.24 & 9.7$\pm$2.6 &  1.72$\pm$0.36  &  6.1  & $-$14.5\\
MATLAS-34$^{(d)}$ & 20.396175 &	3.563550 & 2628 & 9053.4$\pm$36.1 & --- & --- & ---  & 10.3 & $-$15.8  \\
\hline
\end{tabular}
\tablefoot{ Same as Table\,\ref{tab:sample}. (c): MATLAS-23 was not observed with MUSE. Its velocity comes from ALFALFA HI observations \citep{2022ApJS..261...21Y}. (d) the velocity is derived from emissions lines. (10): the distances are estimated from the Hubble flow using a Hubble constant of 70\,km/s/Mpc.}
\label{tab:sample_bg}
\end{table*}

\section{Spatial distribution}
\label{structure}

\begin{figure}[ht]
    \centering
    \includegraphics[width=\linewidth]{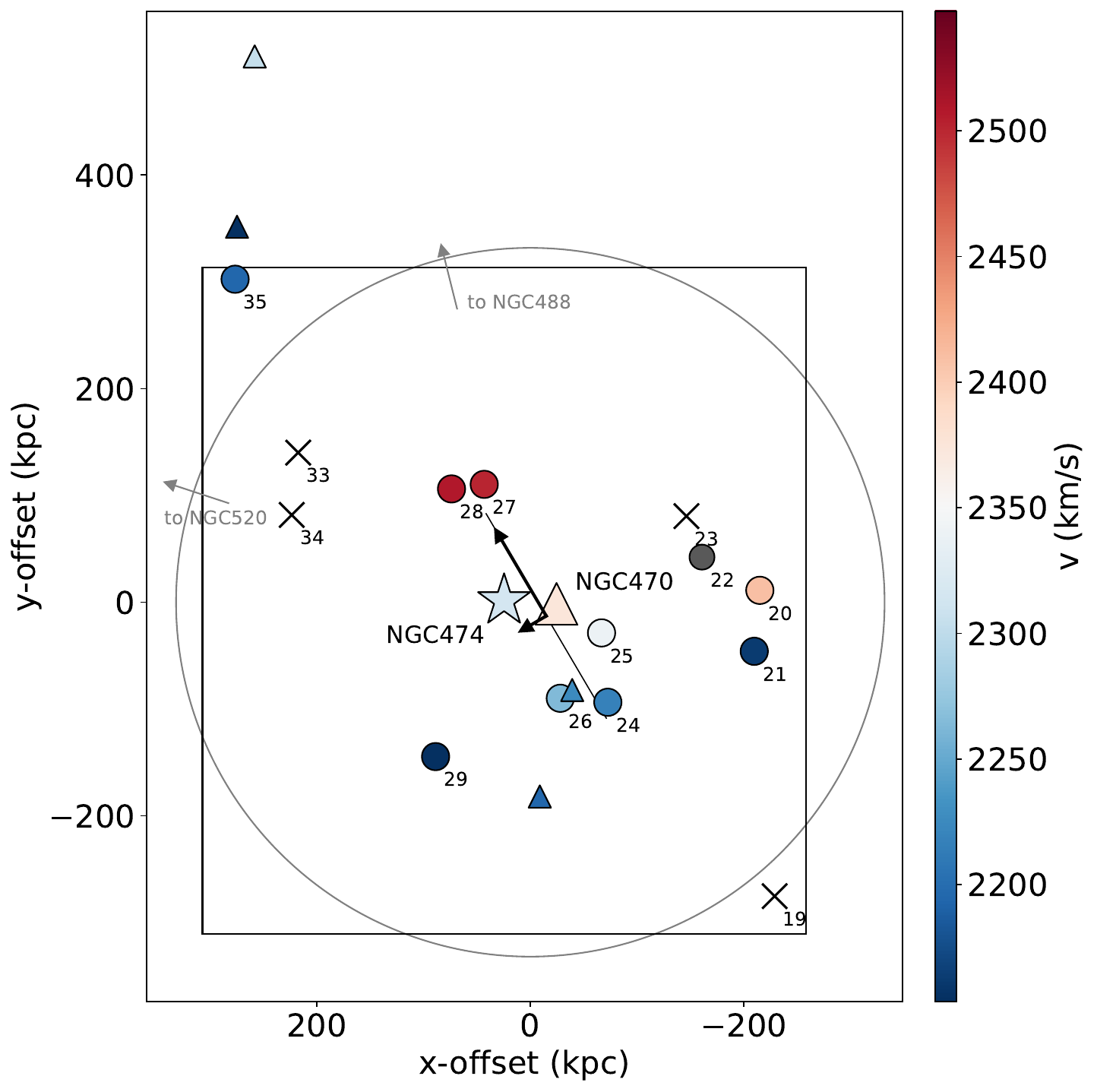}
    \caption{The field in XY coordinates at a distance of 31\,Mpc{, with the field-of-view of the MATLAS survey indicated with the black rectangle}. The X-axis follows the RA-axis, with positive numbers towards East, the Y-axis follows the Dec-axis, with positive numbers towards North. The small numbers indicate the MATLAS index of the dwarf, for example, 20 denotes MATLAS-20.
    Colors indicate the line-of-sight velocities, gray stands for no velocity measurement currently available.  The large star and triangle are NGC\,474 and NGC\,470, respectively. The small triangles are known dwarfs in the field, the dots are the MATLAS dwarfs. Crosses are MATLAS candidates which are rejected as members of the NGC\,474 group based on their velocities.  The large gray circle indicates the virial radius. The line represents the linear substructure, arrows show the minor and major axis and their $rms$ length and thickness. 
    }
    \label{fig:field}
\end{figure}

{
 With the updated list of confirmed and rejected dwarf galaxy members of the NGC\,474 group, we can study the spatial distribution of the satellite system. 
What is the overall flattening considering all dwarf galaxies within the virial radius?
Applying a principal component analysis we recover a semi major axis of 105\,kpc length and a minor-to-major axis ratio of 0.92. This is close to being round and quite different from a flattened distribution. Where is the difference to the previous study coming from?
 }

 {
\citet{2021A&A...654A.161H} uses the Hough transform to find linear structures in data points, while simultaneously maximizing the number of data points in the linear structure and minimizing its thickness. Then they measured the flattening on the sub-population of dwarfs belonging to the linear structure. They found a planar structure with an $rms$ length of 201\,kpc and a $rms$ height of 77\,kpc, consisting of a sub-population of 13 out of the 14 dwarf candidates (effectively all the objects in their catalog belonged to their plane), which extended along the diagonal of the MATLAS footprint. This signal was mainly dominated by two dwarf galaxy candidates (MATLAS-19 and MATLAS-35) in the two opposing corners. Here we find that MATLAS-19 is  actually a background object, and MATLAS-35 lies outside of the edge of the virial radius. The dwarf galaxy MATLAS-35 (top left in Fig.\,\ref{fig:field}) is more distant compared to the other confirmed dwarf galaxies. With an angular separation of 42\,arcmin, it is 380\,kpc from NGC\,474. This is almost the equidistant to NGC\,520 (40\,arcmin, see Fig\,\ref{fig:field_large}). This indicates that MATLAS-35 is not a satellite member, but  resides in the larger field around NGC\,474 (in its vicinity is another dwarf, see Fig\,\ref{fig:field_large}). For the other dwarf which drove the signal -- MATLAS-19 -- we now found based on the MUSE observations that it is not a member of the NGC\,474 group at all, but that it rather is a background galaxy. The updated membership list and restriction to objects within the virial radius therefore changes the estimated flattening of the dwarf galaxy system. However, there might be some interesting correlated sub-population in the data.  
}

{
Applying the Hough transform to find linear structures, we find that six out of the eleven dwarf galaxies and candidate within the virial radius may belong to a linear structure. The center of the linear structure is between NGC\,474 and NGC\,470. The axis ratio is $b/a=0.21$, with the semi minor axis $rms$ length being $b=24$\,kpc and the semi major axis $rms$ length $a=111$\,kpc, and includes MATLAS-24, MATLAS-25, MATLAS-26, MATLAS-27, MATLAS-28, and LEDA\,4765. This is only a sub-population, containing half of the satellites of NGC\,474. Looking at this sub-population, there seems to be a velocity trend (see Fig.\,\ref{fig:field} and next section). We note that for M\,31 roughly half the dwarfs are in a planar structure \citep{2013Natur.493...62I}, however, the absolute number is higher (with 15 out of 27 satellites) and 3D information is available. 
} 

\section{Kinematics}
\label{kinematics}

\begin{figure}[ht]
    \centering
    \includegraphics[width=\linewidth]{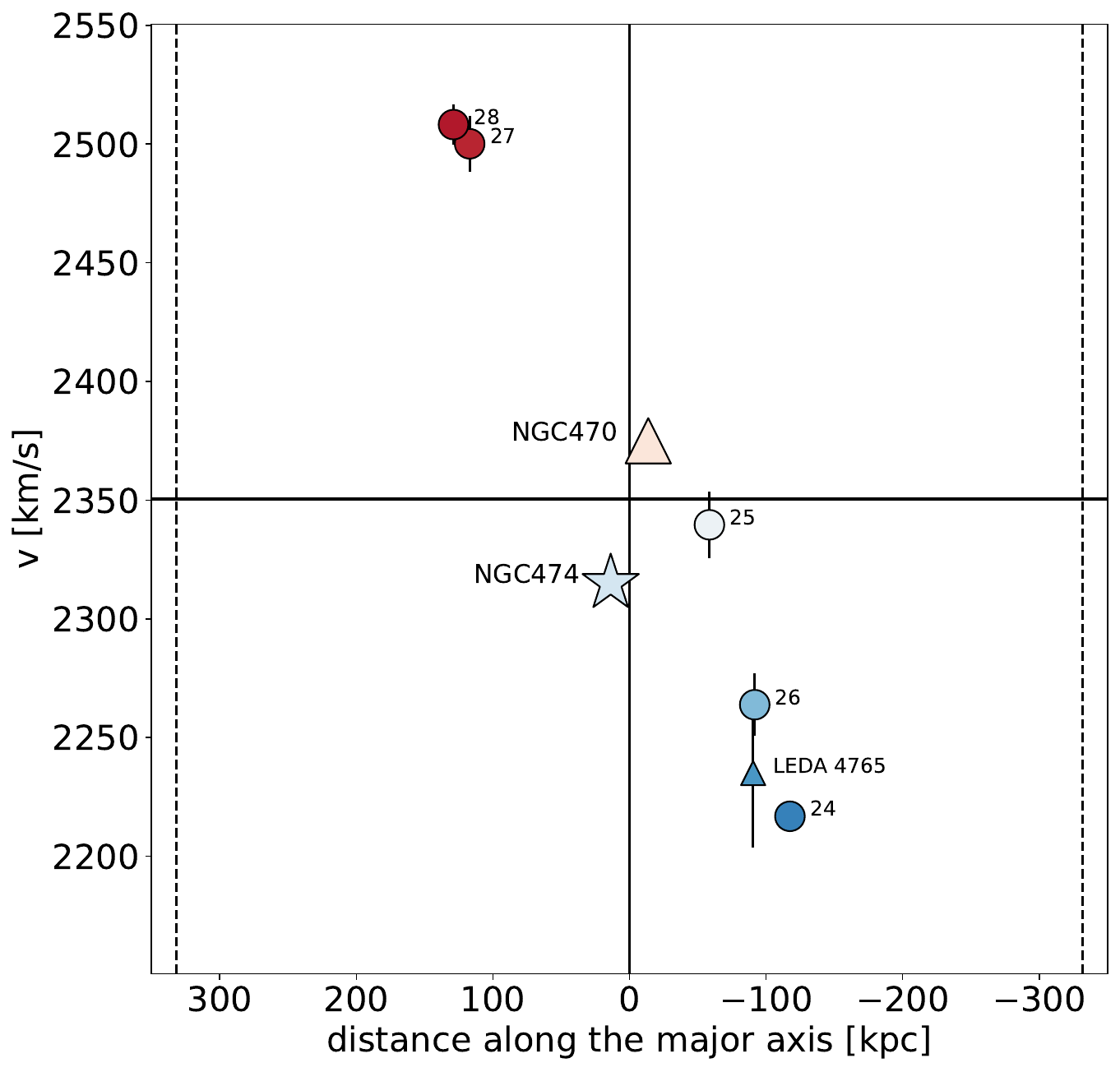}
    \caption{The position velocity diagram. Same point and color scheme as in Fig.\,\ref{fig:field}. The circles indicate dwarfs which are part of the identified linear-structure. Vertical dotted lines correspond to a virial radius of 230\,kpc. The diagram is centered at the  luminosity weighted mean velocity and position of NGC\,474 and NGC\,470.
    }
    \label{fig:pv}
\end{figure}

 There is no clear central galaxy {with both NGC\,474 and NGC\,470 having similar stellar masses}. Therefore, we take the luminosity weighted mean velocity and position as the phase-space center of the group. This follows the approach of \citet{muller2024phase} for the M\,81/M\,82 system. {As discussed in the previous section, there is also no apparent flattening of the full satellite system. However, there seems to a sub-population of six dwarf satellites which seem to be phase-space correlated. Let us quantify this further.}

 If we calculate the distance $\Delta$ along the major axis { of the Hough fit for the sub-population of the six dwarf galaxies}, we can create a position-velocity diagram. As a convention, we assign a negative sign to galaxies occupying the Southern part of the structure. In Fig.\,\ref{fig:pv} we plot the position-velocity diagram, together with the estimated virial radius. {From the six dwarf galaxies five to six occupy two opposing quadrants, with one dwarf -- MATLAS-25 -- being ambiguous due to its uncertainty crossing the quadrants.} {Such a phase-space coherence may arise from co-rotation, however to confirm such an interpretation we would need both tangential and line-of-sight velocities. In general,} if the system were fully pressure supported, all quadrants should be populated equally.  If the system would be fully rotational supported, all satellites should occupy the two opposing quadrants. 
 
 {The probability of finding the observed phase-space correlation  -- i.e. estimating it as a fair coin flip -- is 22\%. This is calculated from the probability of getting 5 or 6 times the same result in a fair coin flip ($p=0.5$), taking into account that we are agnostic on the side of the coin.} It is fairly possible to get this result by chance. {Getting six out of six (instead of at least five out of six) reduces the chance to 3\%. Another way of studying the phase-space correlation of this sub-population is to calculate the Pearson correlation coefficient $r$. This coefficient measures a linear relationship in data points, with values ranging from $-$1 to 1. A Pearson $r$ value close to 1 indicates a strong positive linear correlation, a value close to $-$1 indicates a strong negative linear correlation, and a value close to 0 indicates no linear correlation. For the sub-population of the six satellites, the Pearson $r$ statistic is 0.99, indicating a very strong positive correlation. The $p$-value is 0.00035. Assuming an $\alpha$ level of 0.01 to be statistically significant, this result is indeed significant.}

 The number of satellites for such tests is low, and the signal on one side from NGC\,474 is only driven by two dwarf galaxies – which is likely a pair and do therefore not represent independent tracers. Such pairs can artificially boost the significance. {Further caveats are that we only have access to line-of-sight velocities and projected positions, which may not represent the true 3D velocities and distribution of satellites.}
 The two main galaxies NGC\,474 and NGC\,470 are rotating counter-wise to the satellite system.

\section{Summary and conclusions}
\label{summary}
In this work, we have performed follow up observations with MUSE of the dwarf satellites of the NGC\,474 system, previously discovered in the MATLAS survey. We could confirm nine dwarf galaxies to belong to the group and reject {four} as background objects. One other dwarf galaxy candidate still awaits spectroscopic follow-up.

{Employing full spectrum fitting, we derived velocity and stellar body properties. The dwarfs are in general metal-poor and old. Compared to the Local Group dwarfs, they tend to be of lower metallicity, a finding which was previously reported on 56 MATLAS dwarf galaxies observed with MUSE.}

In a recent work,  \citet{2021A&A...654A.161H} identified the NGC\,474 system to host a significant flattened structure based on the distribution of the dwarf galaxy candidates. With the updated dwarf galaxy catalog, we {reject} their findings {for this system}. {One of the main contributors to the identification of the flattened structure -- MATLAS-19 -- was found to be in the background and  does not belong to the NGC\,474 group, and another -- MATLAS-35 -- is outside of the virial radius. Using a Hough transformation to find linear structures, a flattened system of six dwarf galaxies may still be present in the data. When measuring the flattening of the overall system using its principle components, we find that the axis ratio changes from 0.2 \citep{2021A&A...654A.161H} to 0.9. We therefore conclude that there is no convincing evidence  that all satellites form a plane-of-satellites as previously suggested.}

{However, we cannot rule out the possibility that a sub-population of six dwarf galaxies may form a linear structure around the NGC\,474/NGC\,470 complex. We constructed a position velocity diagram to study any coherent motion within that sub-population.  Out of six dwarf galaxies, five to six seem to coherently move (i.e. being redshifted on one side and blueshifted on the other side). With only six satellites, the signal is not conclusive. However, it is reminiscent of the pair-wise anti-correlated motions of satellites found in SDSS \citep{2014Natur.511..563I}. These have been found to be a challenge for $\Lambda$CDM (\citealt{2015ApJ...805...67I}, but also see \citealt{2015MNRAS.449.2576C}). Interestingly, the orientation of these pair-wise satellite seems to align with the large-scale environment \citep{2014Natur.511..563I}.

Out of the 119 galaxy groups studied in the MATLAS survey with respect to their on-sky distribution, 31 were found to host a statistically significant flattened structure of dwarf galaxy candidates \citep{2021A&A...654A.161H}. Here, we presented the first follow up of one of these significant groups and 
{ find that that previously proposed planar structure {is not} confirmed. Follow-up observations and a more systematic study of such systems would be key to be able to assess the agreement/disagreement with predictions from cosmological models.}

\begin{acknowledgements} 
{We thank the referee for the constructive report, which helped to clarify and improve the manuscript. O.M. thanks Benoit Famaey for interesting discussions concerning the results by \citet{2014Natur.511..563I}.} 
O.M. and N.H. are grateful to the Swiss National Science Foundation for financial support under the grant number  	PZ00P2\_202104. 
M.S.P. acknowledges funding of a Leibniz-Junior Research Group (project number J94/2020). MP is supported by the Academy of Finland grant n:o 347089. S.L. acknowledges the support from the Sejong Science Fellowship Program through the National Research Foundation of Korea (NRF-2021R1C1C2006790).  R.H. acknowledges funding from the Italian INAF Large Grant 12-2022.   S.P. acknowledges support from the Mid-career Researcher Program (No. RS-2023-00208957). This research was supported by the International Space Science Institute (ISSI) in Bern, through ISSI International Team project \#534.

\end{acknowledgements}

\bibliographystyle{aa}
\bibliography{bibliographie}

\begin{thebibliography}{94}
\expandafter\ifx\csname natexlab\endcsname\relax\def\natexlab#1{#1}\fi

\bibitem[{{Bacon} {et~al.}(2010){Bacon}, {Accardo}, {Adjali}, {Anwand}, {Bauer}, {Biswas}, {Blaizot}, {Boudon}, {Brau-Nogue}, {Brinchmann}, {Caillier}, {Capoani}, {Carollo}, {Contini}, {Couderc}, {Daguis{\'e}}, {Deiries}, {Delabre}, {Dreizler}, {Dubois}, {Dupieux}, {Dupuy}, {Emsellem}, {Fechner}, {Fleischmann}, {Fran{\c c}ois}, {Gallou}, {Gharsa}, {Glindemann}, {Gojak}, {Guiderdoni}, {Hansali}, {Hahn}, {Jarno}, {Kelz}, {Koehler}, {Kosmalski}, {Laurent}, {Le Floch}, {Lilly}, {Lizon}, {Loupias}, {Manescau}, {Monstein}, {Nicklas}, {Olaya}, {Pares}, {Pasquini}, {P{\'e}contal-Rousset}, {Pell{\'o}}, {Petit}, {Popow}, {Reiss}, {Remillieux}, {Renault}, {Roth}, {Rupprecht}, {Serre}, {Schaye}, {Soucail}, {Steinmetz}, {Streicher}, {Stuik}, {Valentin}, {Vernet}, {Weilbacher}, {Wisotzki}, \& {Yerle}}]{Bacon2010}
{Bacon}, R., {Accardo}, M., {Adjali}, L., {et~al.} 2010, in \procspie, Vol. 7735, Ground-based and Airborne Instrumentation for Astronomy III, 773508

\bibitem[{{Bacon} {et~al.}(2012){Bacon}, {Accardo}, {Adjali}, {Anwand}, {Bauer}, {Blaizot}, {Boudon}, {Brinchmann}, {Brotons}, {Caillier}, {Capoani}, {Carollo}, {Comin}, {Contini}, {Cumani}, {Daguis}, {Deiries}, {Delabre}, {Dreizler}, {Dubois}, {Dupieux}, {Dupuy}, {Emsellem}, {Fleischmann}, {Fran{\c{c}}ois}, {Gallou}, {Gharsa}, {Girard}, {Glindemann}, {Guiderdoni}, {Hahn}, {Hansali}, {Hofmann}, {Jarno}, {Kelz}, {Kiekebusch}, {Knudstrup}, {Koehler}, {Kollatschny}, {Kosmalski}, {Laurent}, {Le Floch}, {Lilly}, {Lizon {\`a} L'Allemand}, {Loupias}, {Manescau}, {Monstein}, {Nicklas}, {Niemeyer}, {Olaya}, {Palsa}, {Par{\`e}s}, {Pasquini}, {P{\'e}contal-Rousset}, {Pello}, {Petit}, {Piqueras}, {Popow}, {Reiss}, {Remillieux}, {Renault}, {Rhode}, {Richard}, {Roth}, {Rupprecht}, {Schaye}, {Slezak}, {Soucail}, {Steinmetz}, {Streicher}, {Stuik}, {Valentin}, {Vernet}, {Weilbacher}, {Wisotzki}, {Yerle}, \& {Zins}}]{2012Msngr.147....4B}
{Bacon}, R., {Accardo}, M., {Adjali}, L., {et~al.} 2012, The Messenger, 147, 4

\bibitem[{{Barbary}(2016)}]{SEP}
{Barbary}, K. 2016, The Journal of Open Source Software, 1, 58

\bibitem[{{Bertin} \& {Arnouts}(1996)}]{SExtractor}
{Bertin}, E. \& {Arnouts}, S. 1996, \aaps, 117, 393

\bibitem[{{B{\'\i}lek} {et~al.}(2020){B{\'\i}lek}, {Duc}, {Cuillandre}, {Gwyn}, {Cappellari}, {Bekaert}, {Bonfini}, {Bitsakis}, {Paudel}, {Krajnovi{\'c}}, {Durrell}, \& {Marleau}}]{2020MNRAS.498.2138B}
{B{\'\i}lek}, M., {Duc}, P.-A., {Cuillandre}, J.-C., {et~al.} 2020, \mnras, 498, 2138

\bibitem[{{Buck} {et~al.}(2016){Buck}, {Dutton}, \& {Macci{\`o}}}]{2016MNRAS.460.4348B}
{Buck}, T., {Dutton}, A.~A., \& {Macci{\`o}}, A.~V. 2016, \mnras, 460, 4348

\bibitem[{{Buzzo} {et~al.}(2024){Buzzo}, {Forbes}, {Jarrett}, {Marleau}, {Duc}, {Brodie}, {Romanowsky}, {Gannon}, {Janssens}, {Pfeffer}, {Ferr{\'e}-Mateu}, {Haacke}, {Couch}, {Lim}, \& {S{\'a}nchez-Janssen}}]{2024MNRAS.529.3210B}
{Buzzo}, M.~L., {Forbes}, D.~A., {Jarrett}, T.~H., {et~al.} 2024, \mnras, 529, 3210

\bibitem[{{Cantiello} {et~al.}(2007){Cantiello}, {Blakeslee}, {Raimondo}, {Brocato}, \& {Capaccioli}}]{2007ApJ...668..130C}
{Cantiello}, M., {Blakeslee}, J., {Raimondo}, G., {Brocato}, E., \& {Capaccioli}, M. 2007, \apj, 668, 130

\bibitem[{{Cappellari}(2017)}]{2017MNRAS.466..798C}
{Cappellari}, M. 2017, \mnras, 466, 798

\bibitem[{{Cappellari} \& {Emsellem}(2004)}]{2004PASP..116..138C}
{Cappellari}, M. \& {Emsellem}, E. 2004, \pasp, 116, 138

\bibitem[{{Cappellari} {et~al.}(2011){Cappellari}, {Emsellem}, {Krajnovi{\'c}}, {McDermid}, {Scott}, {Verdoes Kleijn}, {Young}, {Alatalo}, {Bacon}, {Blitz}, {Bois}, {Bournaud}, {Bureau}, {Davies}, {Davis}, {de Zeeuw}, {Duc}, {Khochfar}, {Kuntschner}, {Lablanche}, {Morganti}, {Naab}, {Oosterloo}, {Sarzi}, {Serra}, \& {Weijmans}}]{2011MNRAS.413..813C}
{Cappellari}, M., {Emsellem}, E., {Krajnovi{\'c}}, D., {et~al.} 2011, \mnras, 413, 813

\bibitem[{{Cappellari} {et~al.}(2013){Cappellari}, {Scott}, {Alatalo}, {Blitz}, {Bois}, {Bournaud}, {Bureau}, {Crocker}, {Davies}, {Davis}, {de Zeeuw}, {Duc}, {Emsellem}, {Khochfar}, {Krajnovi{\'c}}, {Kuntschner}, {McDermid}, {Morganti}, {Naab}, {Oosterloo}, {Sarzi}, {Serra}, {Weijmans}, \& {Young}}]{2013MNRAS.432.1709C}
{Cappellari}, M., {Scott}, N., {Alatalo}, K., {et~al.} 2013, \mnras, 432, 1709

\bibitem[{{Cautun} {et~al.}(2015{\natexlab{a}}){Cautun}, {Bose}, {Frenk}, {Guo}, {Han}, {Hellwing}, {Sawala}, \& {Wang}}]{2015MNRAS.452.3838C}
{Cautun}, M., {Bose}, S., {Frenk}, C.~S., {et~al.} 2015{\natexlab{a}}, \mnras, 452, 3838

\bibitem[{{Cautun} {et~al.}(2015{\natexlab{b}}){Cautun}, {Wang}, {Frenk}, \& {Sawala}}]{2015MNRAS.449.2576C}
{Cautun}, M., {Wang}, W., {Frenk}, C.~S., \& {Sawala}, T. 2015{\natexlab{b}}, \mnras, 449, 2576

\bibitem[{{Chiboucas} {et~al.}(2013){Chiboucas}, {Jacobs}, {Tully}, \& {Karachentsev}}]{2013AJ....146..126C}
{Chiboucas}, K., {Jacobs}, B.~A., {Tully}, R.~B., \& {Karachentsev}, I.~D. 2013, \aj, 146, 126

\bibitem[{{Conn} {et~al.}(2013){Conn}, {Lewis}, {Ibata}, {Parker}, {Zucker}, {McConnachie}, {Martin}, {Valls-Gabaud}, {Tanvir}, {Irwin}, {Ferguson}, \& {Chapman}}]{2013ApJ...766..120C}
{Conn}, A.~R., {Lewis}, G.~F., {Ibata}, R.~A., {et~al.} 2013, \apj, 766, 120

\bibitem[{{Crosby} {et~al.}(2023){Crosby}, {Jerjen}, {M{\"u}ller}, {Pawlowski}, {Mateo}, \& {Dirnberger}}]{2023MNRAS.521.4009C}
{Crosby}, E., {Jerjen}, H., {M{\"u}ller}, O., {et~al.} 2023, \mnras, 521, 4009

\bibitem[{{Danieli} {et~al.}(2019){Danieli}, {van Dokkum}, {Conroy}, {Abraham}, \& {Romanowsky}}]{2019ApJ...874L..12D}
{Danieli}, S., {van Dokkum}, P., {Conroy}, C., {Abraham}, R., \& {Romanowsky}, A.~J. 2019, \apjl, 874, L12

\bibitem[{{Duc} {et~al.}(2015){Duc}, {Cuillandre}, {Karabal}, {Cappellari}, {Alatalo}, {Blitz}, {Bournaud}, {Bureau}, {Crocker}, {Davies}, {Davis}, {de Zeeuw}, {Emsellem}, {Khochfar}, {Krajnovi{\'c}}, {Kuntschner}, {McDermid}, {Michel-Dansac}, {Morganti}, {Naab}, {Oosterloo}, {Paudel}, {Sarzi}, {Scott}, {Serra}, {Weijmans}, \& {Young}}]{2015MNRAS.446..120D}
{Duc}, P.-A., {Cuillandre}, J.-C., {Karabal}, E., {et~al.} 2015, \mnras, 446, 120

\bibitem[{{Emsellem} {et~al.}(2019){Emsellem}, {van der Burg}, {Fensch}, {Je{\v{r}}{\'a}bkov{\'a}}, {Zanella}, {Agnello}, {Hilker}, {M{\"u}ller}, {Rejkuba}, {Duc}, {Durrell}, {Habas}, {Lelli}, {Lim}, {Marleau}, {Peng}, \& {S{\'a}nchez-Janssen}}]{2019A&A...625A..76E}
{Emsellem}, E., {van der Burg}, R. F.~J., {Fensch}, J., {et~al.} 2019, \aap, 625, A76

\bibitem[{{Fahrion} {et~al.}(2022){Fahrion}, {Bulichi}, {Hilker}, {Leaman}, {Lyubenova}, {M{\"u}ller}, {Neumayer}, {Pinna}, {Rejkuba}, \& {van de Ven}}]{2022A&A...667A.101F}
{Fahrion}, K., {Bulichi}, T.-E., {Hilker}, M., {et~al.} 2022, \aap, 667, A101

\bibitem[{{Fahrion} {et~al.}(2019){Fahrion}, {Lyubenova}, {van de Ven}, {Leaman}, {Hilker}, {Mart{\'\i}n-Navarro}, {Zhu}, {Alfaro-Cuello}, {Coccato}, {Corsini}, {Falc{\'o}n-Barroso}, {Iodice}, {McDermid}, {Sarzi}, \& {de Zeeuw}}]{2019A&A...628A..92F}
{Fahrion}, K., {Lyubenova}, M., {van de Ven}, G., {et~al.} 2019, \aap, 628, A92

\bibitem[{{Fahrion} {et~al.}(2020){Fahrion}, {M{\"u}ller}, {Rejkuba}, {Hilker}, {Lyubenova}, {van de Ven}, {Georgiev}, {Lelli}, {Pawlowski}, \& {Jerjen}}]{2020A&A...634A..53F}
{Fahrion}, K., {M{\"u}ller}, O., {Rejkuba}, M., {et~al.} 2020, \aap, 634, A53

\bibitem[{{Fensch} {et~al.}(2019){Fensch}, {van der Burg}, {Je{\v{r}}{\'a}bkov{\'a}}, {Emsellem}, {Zanella}, {Agnello}, {Hilker}, {M{\"u}ller}, {Rejkuba}, {Duc}, {Durrell}, {Habas}, {Lim}, {Marleau}, {Peng}, \& {S{\'a}nchez Janssen}}]{2019A&A...625A..77F}
{Fensch}, J., {van der Burg}, R. F.~J., {Je{\v{r}}{\'a}bkov{\'a}}, T., {et~al.} 2019, \aap, 625, A77

\bibitem[{{Fernando} {et~al.}(2017){Fernando}, {Arias}, {Guglielmo}, {Lewis}, {Ibata}, \& {Power}}]{2017MNRAS.465..641F}
{Fernando}, N., {Arias}, V., {Guglielmo}, M., {et~al.} 2017, \mnras, 465, 641

\bibitem[{{Ferr{\'e}-Mateu} {et~al.}(2023){Ferr{\'e}-Mateu}, {Gannon}, {Forbes}, {Buzzo}, {Romanowsky}, \& {Brodie}}]{2023MNRAS.526.4735F}
{Ferr{\'e}-Mateu}, A., {Gannon}, J.~S., {Forbes}, D.~A., {et~al.} 2023, \mnras, 526, 4735

\bibitem[{{G{\'a}mez-Mar{\'\i}n} {et~al.}(2024){G{\'a}mez-Mar{\'\i}n}, {Santos-Santos}, {Dom{\'\i}nguez Tenreiro}, {Pedrosa}, {Tissera}, {{\'A}ngeles G{\'o}mez-Flechoso}, \& {Artal}}]{2024arXiv240203288G}
{G{\'a}mez-Mar{\'\i}n}, M., {Santos-Santos}, I., {Dom{\'\i}nguez Tenreiro}, R., {et~al.} 2024, arXiv e-prints, arXiv:2402.03288

\bibitem[{{Geha} {et~al.}(2010){Geha}, {van der Marel}, {Guhathakurta}, {Gilbert}, {Kalirai}, \& {Kirby}}]{2010ApJ...711..361G}
{Geha}, M., {van der Marel}, R.~P., {Guhathakurta}, P., {et~al.} 2010, \apj, 711, 361

\bibitem[{{Gillet} {et~al.}(2015){Gillet}, {Ocvirk}, {Aubert}, {Knebe}, {Libeskind}, {Yepes}, {Gottl{\"o}ber}, \& {Hoffman}}]{2015ApJ...800...34G}
{Gillet}, N., {Ocvirk}, P., {Aubert}, D., {et~al.} 2015, \apj, 800, 34

\bibitem[{{Gu{\'e}rou} {et~al.}(2017){Gu{\'e}rou}, {Krajnovi{\'c}}, {Epinat}, {Contini}, {Emsellem}, {Bouch{\'e}}, {Bacon}, {Michel-Dansac}, {Richard}, {Weilbacher}, {Schaye}, {Marino}, {den Brok}, \& {Erroz-Ferrer}}]{2017A&A...608A...5G}
{Gu{\'e}rou}, A., {Krajnovi{\'c}}, D., {Epinat}, B., {et~al.} 2017, \aap, 608, A5

\bibitem[{{Habas} {et~al.}(2020){Habas}, {Marleau}, {Duc}, {Durrell}, {Paudel}, {Poulain}, {S{\'a}nchez-Janssen}, {Sreejith}, {Ramasawmy}, {Stemock}, {Leach}, {Cuillandre}, {Gwyn}, {Agnello}, {B{\'\i}lek}, {Fensch}, {M{\"u}ller}, {Peng}, \& {van der Burg}}]{2020MNRAS.491.1901H}
{Habas}, R., {Marleau}, F.~R., {Duc}, P.-A., {et~al.} 2020, \mnras, 491, 1901

\bibitem[{{Hanuschik} {et~al.}(2017){Hanuschik}, {Data Processing}, \& {Quality Control Group}}]{Hanuschik2017}
{Hanuschik}, R., {Data Processing}, \& {Quality Control Group}. 2017, in ESO Calibration Workshop: The Second Generation VLT Instruments and Friends, 15

\bibitem[{{Heesters} {et~al.}(2021){Heesters}, {Habas}, {Marleau}, {M{\"u}ller}, {Duc}, {Poulain}, {Durrell}, {S{\'a}nchez-Janssen}, \& {Paudel}}]{2021A&A...654A.161H}
{Heesters}, N., {Habas}, R., {Marleau}, F.~R., {et~al.} 2021, \aap, 654, A161

\bibitem[{{Heesters} {et~al.}(2023){Heesters}, {M{\"u}ller}, {Marleau}, {Duc}, {S{\'a}nchez-Janssen}, {Poulain}, {Habas}, {Lim}, \& {Durrell}}]{2023A&A...676A..33H}
{Heesters}, N., {M{\"u}ller}, O., {Marleau}, F.~R., {et~al.} 2023, \aap, 676, A33

\bibitem[{{Huchra} {et~al.}(1999){Huchra}, {Vogeley}, \& {Geller}}]{1999ApJS..121..287H}
{Huchra}, J.~P., {Vogeley}, M.~S., \& {Geller}, M.~J. 1999, \apjs, 121, 287

\bibitem[{{Ibata} {et~al.}(2014{\natexlab{a}}){Ibata}, {Ibata}, {Famaey}, \& {Lewis}}]{2014Natur.511..563I}
{Ibata}, N.~G., {Ibata}, R.~A., {Famaey}, B., \& {Lewis}, G.~F. 2014{\natexlab{a}}, \nat, 511, 563

\bibitem[{{Ibata} {et~al.}(2015){Ibata}, {Famaey}, {Lewis}, {Ibata}, \& {Martin}}]{2015ApJ...805...67I}
{Ibata}, R.~A., {Famaey}, B., {Lewis}, G.~F., {Ibata}, N.~G., \& {Martin}, N. 2015, \apj, 805, 67

\bibitem[{{Ibata} {et~al.}(2014{\natexlab{b}}){Ibata}, {Ibata}, {Lewis}, {Martin}, {Conn}, {Elahi}, {Arias}, \& {Fernando}}]{2014ApJ...784L...6I}
{Ibata}, R.~A., {Ibata}, N.~G., {Lewis}, G.~F., {et~al.} 2014{\natexlab{b}}, \apjl, 784, L6

\bibitem[{{Ibata} {et~al.}(2013){Ibata}, {Lewis}, {Conn}, {Irwin}, {McConnachie}, {Chapman}, {Collins}, {Fardal}, {Ferguson}, {Ibata}, {Mackey}, {Martin}, {Navarro}, {Rich}, {Valls-Gabaud}, \& {Widrow}}]{2013Natur.493...62I}
{Ibata}, R.~A., {Lewis}, G.~F., {Conn}, A.~R., {et~al.} 2013, NAT, 493, 62

\bibitem[{{J{\'u}lio} {et~al.}(2024){J{\'u}lio}, {Pawlowski}, {Sohn}, {Taibi}, {van der Marel}, \& {McGaugh}}]{2024arXiv240416110J}
{J{\'u}lio}, M.~P., {Pawlowski}, M.~S., {Sohn}, S.~T., {et~al.} 2024, arXiv e-prints, arXiv:2404.16110

\bibitem[{{Kanehisa} {et~al.}(2023){Kanehisa}, {Pawlowski}, {M{\"u}ller}, \& {Sohn}}]{2023MNRAS.519.6184K}
{Kanehisa}, K.~J., {Pawlowski}, M.~S., {M{\"u}ller}, O., \& {Sohn}, S.~T. 2023, \mnras, 519, 6184

\bibitem[{{Karachentsev} \& {Kroupa}(2024)}]{2024MNRAS.528.2805K}
{Karachentsev}, I.~D. \& {Kroupa}, P. 2024, \mnras, 528, 2805

\bibitem[{{Kirby} {et~al.}(2013){Kirby}, {Cohen}, {Guhathakurta}, {Cheng}, {Bullock}, \& {Gallazzi}}]{2013ApJ...779..102K}
{Kirby}, E.~N., {Cohen}, J.~G., {Guhathakurta}, P., {et~al.} 2013, \apj, 779, 102

\bibitem[{{Koch} \& {Grebel}(2006)}]{2006AJ....131.1405K}
{Koch}, A. \& {Grebel}, E.~K. 2006, \aj, 131, 1405

\bibitem[{{Kravtsov}(2013)}]{2013ApJ...764L..31K}
{Kravtsov}, A.~V. 2013, \apjl, 764, L31

\bibitem[{{Kroupa}(2001)}]{2001MNRAS.322..231K}
{Kroupa}, P. 2001, \mnras, 322, 231

\bibitem[{{Kroupa} {et~al.}(2005){Kroupa}, {Theis}, \& {Boily}}]{2005A&A...431..517K}
{Kroupa}, P., {Theis}, C., \& {Boily}, C.~M. 2005, \aap, 431, 517

\bibitem[{{Libeskind} {et~al.}(2019){Libeskind}, {Carlesi}, {M{\"u}ller}, {Pawlowski}, {Hoffman}, {Pomar{\`e}de}, {Courtois}, {Tully}, {Gottl{\"o}ber}, {Steinmetz}, {Sorce}, \& {Knebe}}]{2019MNRAS.490.3786L}
{Libeskind}, N.~I., {Carlesi}, E., {M{\"u}ller}, O., {et~al.} 2019, \mnras, 490, 3786

\bibitem[{{Libeskind} {et~al.}(2007){Libeskind}, {Cole}, {Frenk}, {Okamoto}, \& {Jenkins}}]{2007MNRAS.374...16L}
{Libeskind}, N.~I., {Cole}, S., {Frenk}, C.~S., {Okamoto}, T., \& {Jenkins}, A. 2007, \mnras, 374, 16

\bibitem[{{Libeskind} {et~al.}(2015){Libeskind}, {Hoffman}, {Tully}, {Courtois}, {Pomar{\`e}de}, {Gottl{\"o}ber}, \& {Steinmetz}}]{2015MNRAS.452.1052L}
{Libeskind}, N.~I., {Hoffman}, Y., {Tully}, R.~B., {et~al.} 2015, \mnras, 452, 1052

\bibitem[{{Lu} {et~al.}(1993){Lu}, {Hoffman}, {Groff}, {Roos}, \& {Lamphier}}]{1993ApJS...88..383L}
{Lu}, N.~Y., {Hoffman}, G.~L., {Groff}, T., {Roos}, T., \& {Lamphier}, C. 1993, \apjs, 88, 383

\bibitem[{{Lynden-Bell}(1976)}]{1976MNRAS.174..695L}
{Lynden-Bell}, D. 1976, \mnras, 174, 695

\bibitem[{{Mart{\'\i}nez-Delgado} {et~al.}(2021){Mart{\'\i}nez-Delgado}, {Makarov}, {Javanmardi}, {Pawlowski}, {Makarova}, {Donatiello}, {Lang}, {Rom{\'a}n}, {Vivas}, \& {Carballo-Bello}}]{2021A&A...652A..48M}
{Mart{\'\i}nez-Delgado}, D., {Makarov}, D., {Javanmardi}, B., {et~al.} 2021, \aap, 652, A48

\bibitem[{{McConnachie}(2012)}]{2012AJ....144....4M}
{McConnachie}, A.~W. 2012, AJ, 144, 4

\bibitem[{{McConnachie} \& {Irwin}(2006)}]{2006MNRAS.365..902M}
{McConnachie}, A.~W. \& {Irwin}, M.~J. 2006, \mnras, 365, 902

\bibitem[{{McGaugh}(2012)}]{2012AJ....143...40M}
{McGaugh}, S.~S. 2012, \aj, 143, 40

\bibitem[{{Metz} {et~al.}(2008){Metz}, {Kroupa}, \& {Libeskind}}]{2008ApJ...680..287M}
{Metz}, M., {Kroupa}, P., \& {Libeskind}, N.~I. 2008, \apj, 680, 287

\bibitem[{{M{\"u}ller} {et~al.}(2021{\natexlab{a}}){M{\"u}ller}, {Fahrion}, {Rejkuba}, {Hilker}, {Lelli}, {Lutz}, {Pawlowski}, {Coccato}, {Anand}, \& {Jerjen}}]{Muller2021a}
{M{\"u}ller}, O., {Fahrion}, K., {Rejkuba}, M., {et~al.} 2021{\natexlab{a}}, \aap, 645, A92

\bibitem[{{M{\"u}ller} {et~al.}(2024){M{\"u}ller}, {Heesters}, {Pawlowski}, {Jamie Kanehisa}, {Lelli}, \& {Libeskind}}]{2024A&A...683A.250M}
{M{\"u}ller}, O., {Heesters}, N., {Pawlowski}, M.~S., {et~al.} 2024, \aap, 683, A250

\bibitem[{M{\"u}ller {et~al.}(2024)M{\"u}ller, Heesters, Pawlowski, Kanehisa, Lelli, \& Libeskind}]{muller2024phase}
M{\"u}ller, O., Heesters, N., Pawlowski, M.~S., {et~al.} 2024, Astronomy \& Astrophysics, 683, A250

\bibitem[{{M{\"u}ller} {et~al.}(2016){M{\"u}ller}, {Jerjen}, {Pawlowski}, \& {Binggeli}}]{Muller2016}
{M{\"u}ller}, O., {Jerjen}, H., {Pawlowski}, M.~S., \& {Binggeli}, B. 2016, \aap, 595, A119

\bibitem[{{M{\"u}ller} {et~al.}(2020){M{\"u}ller}, {Marleau}, {Duc}, {Habas}, {Fensch}, {Emsellem}, {Poulain}, {Lim}, {Agnello}, {Durrell}, {Paudel}, {S{\'a}nchez-Janssen}, \& {van der Burg}}]{2020A&A...640A.106M}
{M{\"u}ller}, O., {Marleau}, F.~R., {Duc}, P.-A., {et~al.} 2020, \aap, 640, A106

\bibitem[{{M{\"u}ller} {et~al.}(2018{\natexlab{a}}){M{\"u}ller}, {Pawlowski}, {Jerjen}, \& {Lelli}}]{2018Sci...359..534M}
{M{\"u}ller}, O., {Pawlowski}, M.~S., {Jerjen}, H., \& {Lelli}, F. 2018{\natexlab{a}}, Science, 359, 534

\bibitem[{{M{\"u}ller} {et~al.}(2021{\natexlab{b}}){M{\"u}ller}, {Pawlowski}, {Lelli}, {Fahrion}, {Rejkuba}, {Hilker}, {Kanehisa}, {Libeskind}, \& {Jerjen}}]{Muller2021b}
{M{\"u}ller}, O., {Pawlowski}, M.~S., {Lelli}, F., {et~al.} 2021{\natexlab{b}}, \aap, 645, L5

\bibitem[{{M{\"u}ller} {et~al.}(2018{\natexlab{b}}){M{\"u}ller}, {Rejkuba}, \& {Jerjen}}]{MuellerTRGB2018}
{M{\"u}ller}, O., {Rejkuba}, M., \& {Jerjen}, H. 2018{\natexlab{b}}, \aap, 615, A96

\bibitem[{{M{\"u}ller} {et~al.}(2019){M{\"u}ller}, {Rejkuba}, {Pawlowski}, {Ibata}, {Lelli}, {Hilker}, \& {Jerjen}}]{Muller2019}
{M{\"u}ller}, O., {Rejkuba}, M., {Pawlowski}, M.~S., {et~al.} 2019, \aap, 629, A18

\bibitem[{{M{\"u}ller} {et~al.}(2017){M{\"u}ller}, {Scalera}, {Binggeli}, \& {Jerjen}}]{2017A&A...602A.119M}
{M{\"u}ller}, O., {Scalera}, R., {Binggeli}, B., \& {Jerjen}, H. 2017, \aap, 602, A119

\bibitem[{{Mutlu-Pakdil} {et~al.}(2024){Mutlu-Pakdil}, {Sand}, {Crnojevi{\'c}}, {Bennet}, {Jones}, {Spekkens}, {Karunakaran}, {Zaritsky}, {Caldwell}, {Fielder}, {Guhathakurta}, {Seth}, {Simon}, {Strader}, \& {Toloba}}]{2024ApJ...966..188M}
{Mutlu-Pakdil}, B., {Sand}, D.~J., {Crnojevi{\'c}}, D., {et~al.} 2024, \apj, 966, 188

\bibitem[{{Paudel} {et~al.}(2021){Paudel}, {Yoon}, \& {Smith}}]{2021ApJ...917L..18P}
{Paudel}, S., {Yoon}, S.-J., \& {Smith}, R. 2021, \apjl, 917, L18

\bibitem[{{Pawlowski}(2018)}]{2018MPLA...3330004P}
{Pawlowski}, M.~S. 2018, Modern Physics Letters A, 33, 1830004

\bibitem[{{Pawlowski}(2021)}]{2021NatAs...5.1185P}
{Pawlowski}, M.~S. 2021, Nature Astronomy, 5, 1185

\bibitem[{{Pawlowski} \& {Kroupa}(2013)}]{2013MNRAS.435.2116P}
{Pawlowski}, M.~S. \& {Kroupa}, P. 2013, \mnras, 435, 2116

\bibitem[{{Pawlowski} {et~al.}(2024){Pawlowski}, {M{\"u}ller}, {Taibi}, {J{\'u}lio}, {Kanehisa}, \& {Heesters}}]{2024arXiv240506016P}
{Pawlowski}, M.~S., {M{\"u}ller}, O., {Taibi}, S., {et~al.} 2024, arXiv e-prints, arXiv:2405.06016

\bibitem[{{Pawlowski} {et~al.}(2012){Pawlowski}, {Pflamm-Altenburg}, \& {Kroupa}}]{2012MNRAS.423.1109P}
{Pawlowski}, M.~S., {Pflamm-Altenburg}, J., \& {Kroupa}, P. 2012, \mnras, 423, 1109

\bibitem[{{Poulain} {et~al.}(2021){Poulain}, {Marleau}, {Habas}, {Duc}, {S{\'a}nchez-Janssen}, {Durrell}, {Paudel}, {Ahad}, {Chougule}, {M{\"u}ller}, {Lim}, {B{\'\i}lek}, \& {Fensch}}]{2021MNRAS.506.5494P}
{Poulain}, M., {Marleau}, F.~R., {Habas}, R., {et~al.} 2021, \mnras, 506, 5494

\bibitem[{{Rines} {et~al.}(2003){Rines}, {Geller}, {Kurtz}, \& {Diaferio}}]{2003AJ....126.2152R}
{Rines}, K., {Geller}, M.~J., {Kurtz}, M.~J., \& {Diaferio}, A. 2003, \aj, 126, 2152

\bibitem[{{Sales} {et~al.}(2022){Sales}, {Wetzel}, \& {Fattahi}}]{2022NatAs...6..897S}
{Sales}, L.~V., {Wetzel}, A., \& {Fattahi}, A. 2022, Nature Astronomy, 6, 897

\bibitem[{{Santos-Santos} {et~al.}(2023){Santos-Santos}, {G{\'a}mez-Mar{\'\i}n}, {Dom{\'\i}nguez-Tenreiro}, {Tissera}, {Bignone}, {Pedrosa}, {Artal}, {G{\'o}mez-Flechoso}, {Rufo-Pastor}, {Mart{\'\i}nez-Serrano}, \& {Serna}}]{2023ApJ...942...78S}
{Santos-Santos}, I., {G{\'a}mez-Mar{\'\i}n}, M., {Dom{\'\i}nguez-Tenreiro}, R., {et~al.} 2023, \apj, 942, 78

\bibitem[{{Sato} \& {Chiba}(2024)}]{2024arXiv240108143S}
{Sato}, G. \& {Chiba}, M. 2024, arXiv e-prints, arXiv:2401.08143

\bibitem[{{Schlafly} \& {Finkbeiner}(2011)}]{2011ApJ...737..103S}
{Schlafly}, E.~F. \& {Finkbeiner}, D.~P. 2011, \apj, 737, 103

\bibitem[{{Sheth} {et~al.}(2010){Sheth}, {Regan}, {Hinz}, {Gil de Paz}, {Men{\'e}ndez-Delmestre}, {Mu{\~n}oz-Mateos}, {Seibert}, {Kim}, {Laurikainen}, {Salo}, {Gadotti}, {Laine}, {Mizusawa}, {Armus}, {Athanassoula}, {Bosma}, {Buta}, {Capak}, {Jarrett}, {Elmegreen}, {Elmegreen}, {Knapen}, {Koda}, {Helou}, {Ho}, {Madore}, {Masters}, {Mobasher}, {Ogle}, {Peng}, {Schinnerer}, {Surace}, {Zaritsky}, {Comer{\'o}n}, {de Swardt}, {Meidt}, {Kasliwal}, \& {Aravena}}]{2010PASP..122.1397S}
{Sheth}, K., {Regan}, M., {Hinz}, J.~L., {et~al.} 2010, \pasp, 122, 1397

\bibitem[{{Sohn} {et~al.}(2020){Sohn}, {Patel}, {Fardal}, {Besla}, {van der Marel}, {Geha}, \& {Guhathakurta}}]{2020ApJ...901...43S}
{Sohn}, S.~T., {Patel}, E., {Fardal}, M.~A., {et~al.} 2020, \apj, 901, 43

\bibitem[{{Soto} {et~al.}(2016){Soto}, {Lilly}, {Bacon}, {Richard}, \& {Conseil}}]{Soto2016}
{Soto}, K.~T., {Lilly}, S.~J., {Bacon}, R., {Richard}, J., \& {Conseil}, S. 2016, \mnras, 458, 3210

\bibitem[{{Springob} {et~al.}(2005){Springob}, {Haynes}, {Giovanelli}, \& {Kent}}]{2005ApJS..160..149S}
{Springob}, C.~M., {Haynes}, M.~P., {Giovanelli}, R., \& {Kent}, B.~R. 2005, \apjs, 160, 149

\bibitem[{{Taibi} {et~al.}(2024){Taibi}, {Pawlowski}, {Khoperskov}, {Steinmetz}, \& {Libeskind}}]{2024A&A...681A..73T}
{Taibi}, S., {Pawlowski}, M.~S., {Khoperskov}, S., {Steinmetz}, M., \& {Libeskind}, N.~I. 2024, \aap, 681, A73

\bibitem[{Teeninga {et~al.}(2015)Teeninga, Moschini, Trager, \& Wilkinson}]{teeninga2015improved}
Teeninga, P., Moschini, U., Trager, S.~C., \& Wilkinson, M.~H. 2015, in International Symposium on Mathematical Morphology and Its Applications to Signal and Image Processing, Springer, 157--168

\bibitem[{{Tully} {et~al.}(2013){Tully}, {Courtois}, {Dolphin}, {Fisher}, {H{\'e}raudeau}, {Jacobs}, {Karachentsev}, {Makarov}, {Makarova}, {Mitronova}, {Rizzi}, {Shaya}, {Sorce}, \& {Wu}}]{2013AJ....146...86T}
{Tully}, R.~B., {Courtois}, H.~M., {Dolphin}, A.~E., {et~al.} 2013, \aj, 146, 86

\bibitem[{{Tully} {et~al.}(2015){Tully}, {Libeskind}, {Karachentsev}, {Karachentseva}, {Rizzi}, \& {Shaya}}]{2015ApJ...802L..25T}
{Tully}, R.~B., {Libeskind}, N.~I., {Karachentsev}, I.~D., {et~al.} 2015, \apjl, 802, L25

\bibitem[{{Vazdekis} {et~al.}(2016){Vazdekis}, {Koleva}, {Ricciardelli}, {R{\"o}ck}, \& {Falc{\'o}n-Barroso}}]{2016MNRAS.463.3409V}
{Vazdekis}, A., {Koleva}, M., {Ricciardelli}, E., {R{\"o}ck}, B., \& {Falc{\'o}n-Barroso}, J. 2016, \mnras, 463, 3409

\bibitem[{{Wang} {et~al.}(2013){Wang}, {Frenk}, \& {Cooper}}]{2013MNRAS.429.1502W}
{Wang}, J., {Frenk}, C.~S., \& {Cooper}, A.~P. 2013, \mnras, 429, 1502

\bibitem[{{Wang} {et~al.}(2018){Wang}, {Luo}, {Shen}, {Hou}, {Kong}, {Song}, {Zhang}, {Wu}, {Cao}, {Hou}, {Wang}, {Zhang}, \& {Zhao}}]{2018MNRAS.474.1873W}
{Wang}, L.-L., {Luo}, A.~L., {Shen}, S.-Y., {et~al.} 2018, \mnras, 474, 1873

\bibitem[{{Weilbacher} {et~al.}(2020){Weilbacher}, {Palsa}, {Streicher}, {Bacon}, {Urrutia}, {Wisotzki}, {Conseil}, {Husemann}, {Jarno}, {Kelz}, {P{\'e}contal-Rousset}, {Richard}, {Roth}, {Selman}, \& {Vernet}}]{2020arXiv200608638W}
{Weilbacher}, P.~M., {Palsa}, R., {Streicher}, O., {et~al.} 2020, arXiv e-prints, arXiv:2006.08638

\bibitem[{{Weilbacher} {et~al.}(2012){Weilbacher}, {Streicher}, {Urrutia}, {Jarno}, {P{\'e}contal-Rousset}, {Bacon}, \& {B{\"o}hm}}]{Weilbacher2012}
{Weilbacher}, P.~M., {Streicher}, O., {Urrutia}, T., {et~al.} 2012, in \procspie, Vol. 8451, Software and Cyberinfrastructure for Astronomy II, 84510B

\bibitem[{{Yu} {et~al.}(2022){Yu}, {Ho}, {Wang}, \& {Li}}]{2022ApJS..261...21Y}
{Yu}, N., {Ho}, L.~C., {Wang}, J., \& {Li}, H. 2022, \apjs, 261, 21

\end{thebibliography}

\end{document}